# Mixed-Valence Tetranuclear Manganese Single-Molecule Magnets

Jae Yoo,[1a] Akira Yamaguchi,[1b] Motohiro Nakano,[1a] J. Krzystek,[1c] William E. Streib,[1d] Louis-Claude Brunel,[1c] Hidehiko Ishimoto,[1b] George Christou,*[1a] and David N. Hendrickson*[1a]

Department of Chemistry and Biochemistry - 0358, University of California at San Diego, La Jolla, CA  92093; Institute of Solid State Physics, The University of Tokyo, 7-22-1 Roppongi, Minatoku, Tokyo 106-8666, Japan; Center for Interdisciplinary Magnetic Resonance, National High Magnetic Field Laboratory, Tallahassee, FL  32310; and Department of Chemistry and Molecular Structure Center, Indiana University, Bloomington, IN  47405.

**Abstract**

The preparations, X-ray structures, and detailed physical characterizations are presented for two new mixed-valence tetranuclear manganese complexes that function as single-molecule magnets (SMM's): $[Mn_4(hmp)_6Br_2(H_2O)_2]Br_2$ **(2)** and $[Mn_4(6me-hmp)_6Cl_4]\cdot 4H_2O$ **(3)**, where hmpH is 2-hydroxymethylpyridine and 6me-hmp is 6-methyl-2-hydroxymethylpyridine. Complex **2**·$H_2O$ crystallizes in the $P2_1/c$, with cell dimensions at -160 °C of $a$ = 10.907(0) Å, $b$ = 15.788(0) Å, $c$ = 13.941(0) Å, $\beta$ = 101.21(0)°, and $Z$ = 2.  The cation lies on an inversion center and consists of a planar $Mn_4$ rhombus that is mixed-valent, $Mn_2^{III}Mn_2^{II}$.  The hmp$^-$ ligands function as bidentate ligands, and as the only bridging ligand in **2**·4$H_2O$.  Complex

**3·4H$_2$O** crystallizes in the monoclinic space group *C*2/c, with cell dimensions at -160 °C of *a* = 17.0852(4) Å, *b* = 20.8781(5) Å, *c* = 14.835(3) Å, *β* = 90.5485(8)°, and *Z* = 4. This neutral complex has a mixed-valence Mn$_2^{III}$Mn$_2^{II}$ composition and is best described as having four manganese ions arranged in a bent chain. An µ$_2$-oxygen atom of the 6me-hmp$^-$ anion bridges between the manganese ions; the Cl$^-$ ligands are terminal. Variable-field magnetization and High-frequency and -field EPR (HFEPR) data indicate that complexes **2·4H$_2$O** and **3·4H$_2$O** have *S* = 9 ground states. Fine structure patterns are seen in the HFEPR (191.30, 291.26, 388.35 GHz) spectra, and in the case of **2·4H$_2$O**, it was possible to simulate the fine structure assuming *S* = 9 with the parameters *g* = 1.999, axial zero-field splitting of *D*/*k*$_B$ = -0.498 K, and quartic longitudinal zero-field splitting of *B*$_4^0$/*k*$_B$ = 1.72 x 10$^{-5}$ K. Complex **2·4H$_2$O** exhibits a frequency-dependent out-of-phase AC magnetic susceptibility signal, clearly indicating that this complex functions as a SMM. The AC susceptibility data for complex **2·4H$_2$O** were measured in the 0.05 - 4.0 K range and when fit to the Arrhenius law, gave an activation energy of Δ*E* = 15.8 K for the reversal of magnetization. This Δ*E* value is to be compared to the potential-energy barrier height of $U = |D\hat{S}_z^2|$ = 40.3 K calculated for **2·4H$_2$O**.

**Introduction**

    Single-molecule magnets (SMM's) are of considerable interest because they could serve as small magnetic memory units.[2,3] SMM's are also being studied since they provide a means to systematically study the chemistry and physics of nanomagnets.[3] They can be viewed as slowly

relaxing magnetic particles of well-defined size and shape. The slow magnetization relaxation of a SMM is due to the presence of an energy barrier for the reversal of the direction of its magnetization. The magnetization of a molecule that functions as a SMM can be either aligned "spin-up" or spin-down" relative to the principal axis of magnetization of the molecule. In simple terms, such a molecule has a large spin ($S$) ground state and the barrier for flipping the magnetic moment from "spin-up" to "spin-down" is due to a Ising type magnetic anisotropy characterized by, $(D\hat{S}_Z^2)$, where $D < 0$. The potential energy diagram for a $S = 9$ SMM is shown in Figure 1, where the minimum with $M_S = -9$ on the left corresponds to the "spin-up" state and the minimum ($M_S = +9$) on the right of the energy barrier is the "spin-down" state. The height of the energy barrier is given by $U = |D\hat{S}^2|$. The origin of the magnetic anisotropy $(D\hat{S}_Z^2)$ is single-ion zero-field splittings at the Mn atom sites in the molecule. At low temperatures, the conversion from "spin-up" to "spin-down" becomes sluggish and this results in the appearance of an out-of-phase AC magnetic susceptibility signal and magnetization hysteresis loops.

The first molecule shown to be a SMM is [Mn$_{12}$O$_{12}$(O$_2$CMe)$_{16}$(H$_2$O)$_4$]·4H$_2$O·2HO$_2$Cme[4], named Mn12Ac in the literature,[5-51] which exhibits slow magnetization relaxation due to a $S = 10$ ground state split by axial zero-field splitting $(D\hat{S}_Z^2)$, where $D/k_B = -0.61$ K. In 1996, Mn12Ac was reported[17-20] to show quantum tunneling of magnetization. Steps at regular intervals of magnetic field were seen in the magnetization hysteresis loops for oriented crystals of Mn12Ac and this generated much interest in its magnetic properties.

An important research goal in the field of SMM's has been the discovery of new examples of SMM's. This can lead to the understanding of the factors that govern the relaxation process. In addition to Mn12Ac, structurally related dodecanuclear manganese SMM's with the composition [Mn$_{12}$O$_{12}$(O$_2$CR)$_{16}$(H$_2$O)$_4$][52-61] and their anionic derivatives, [cation][Mn$_{12}$O$_{12}$(O$_2$CR)$_{16}$(H$_2$O)$_4$],[62,63] have been characterized. Quantum tunneling of magnetization has been observed for not only the integer spin ground state ($S$ = 10) Mn$_{12}$ complexes, but also for the anionic [Mn$_{12}^-$] derivatives with half-odd integer spin ground state ($S$ = 19/2). Several families of tetranuclear SMM's have been discovered as well: a tetranuclear vanadium complex,[64] several manganese complexes with a distorted cubane core [Mn$^{IV}$Mn$^{III}_3$O$_3$X]$^{6+}$,[65,66] and an iron complex of the form [Fe$_4$(OCH$_3$)$_6$(dpm)$_6$] where dpm is the anion of dipivaloylmethane.[67] Finally, an octanuclear iron complex, [Fe$_8$O$_2$(OH)$_{12}$(tacn)$_6$]$^{8+}$, where tacn is triazacyclononane has also been characterized as an SMM.[68]

In our efforts to find new examples of SMM's, we very recently characterized a new manganese based SMM with the chemical composition [Mn$_4$(pdmH)$_6$(O$_2$CCH$_3$)$_2$](ClO$_4$)$_2$ (**1**)[69] incorporating the ligand pyridine dimethanol (pdmH$^-$). The magnetic properties of this complex were fully characterized by DC and AC magnetic susceptibility and by high-frequency EPR (HFEPR) measurements. The hydrated form of complex **1** (**1**·2.5H$_2$O) was determined by HFEPR to have a ground state spin of $S$ = 9 and axial zero-field splitting parameter of $D/k_B$ = -0.45 K giving it a barrier for the reversal of magnetization of 36.5 K. Frequency dependent out-of-phase AC magnetic susceptibility signals in the 2 K region were seen. A temperature-independent magnetization relaxation time on the order of 10$^4$ seconds was observed below 1

K.[70] This was attributed to quantum mechanical tunneling of the direction of magnetization between the $M_S$ = -9 and $M_S$ = +9 lowest components of the $S$ = 9 ground state. This is significant observation that adds to our understanding of magnetization relaxation in SMM's.

We herein present results on our latest efforts in the discovery of new SMM's. Two new tetranuclear mixed-valence manganese complexes were isolated using ligands similar to pdmH⁻ (see Figure 2). With hmpH, a complex of the form [Mn$_4$(hmp)$_6$Br$_2$(H$_2$O)$_2$]Br$_2$ (**2**) was isolated whereas with 6me-hmpH, a structurally different complex of the form [Mn$_4$(6me-hmp)$_6$Cl$_4$] (**3**) was found. The synthesis, X-ray structure, and magnetic characterization for complexes **2** and **3** are presented.

**Experimental Section**

**Synthesis**. All manipulations were carried out under aerobic conditions using commercial grade solvents. Manganous chloride, manganous bromide, 2-hydroxymethyl-pyridine (pyridyl-carbinol, Aldrich), 6-methyl-2-hydroxymethylpyridine (Aldrich), and tetraethylammonium hydroxide were all used as received without further purification. Tetrabutylammonium permanganate was prepared as described elsewhere.[71]

**[Mn$_4$(hmp)$_6$Br$_2$(H$_2$O)$_2$]Br$_2$ (2)**. 376 mg of 2-hydroxymethylpyridine (hmpH) and 395 mg of MnBr$_2$·4H$_2$O were dissolved in 20 mL of MeCN. To this was added 530 mg of a 20% wt water solution of tetraethylammonium hydroxide. The solution turned red-brown and this was stirred for 40-60 minutes and then filtered. The filtrate was then left to slowly evaporate. Dark cube-like crystals formed in 1-2 weeks. Anal. Calc (Found) for C$_{36}$H$_{48}$Br$_4$Mn$_4$N$_6$O$_{12}$: C, 33.36 (33.83); H, 3.73 (3.47); N, 6.48 (6.56). Selected IR data (cm$^{-1}$): 3350.2 (br), 1604.7 (s),

1483.2 (m), 1438.8 (m), 1282.6 (w), 1226.7 (w), 1155.3 (w), 1045.4 (s), 765.7 (s), 659.6 (s), 567.1 (s).

**[Mn$_4$(6me-hmp)$_6$Cl$_4$]·4H$_2$O (3)**. 432 mg of 6-methyl-2-hydroxymethylpyridine (6me-hmp) and 273 mg of MnCl$_2$·4H$_2$O were dissolved in 20 mL of MeOH. To this was added in portions 125 mg of tetrabutylammonium permanganate. The solution turned an intense brown color. This solution was stirred for one hour. An equal volume of MeCN was added and the mixture was left to slowly evaporate. Dark prism-shaped crystals formed over several days. Anal. Calc (Found) for C$_{42}$H$_{56}$Cl$_4$Mn$_4$N$_6$O$_{10}$: C, 43.25 (43.26); H, 4.84 (4.70); N, 7.20 (7.43). Selected IR data (cm$^{-1}$): 3506.4 (br), 1606.6 (m), 1465.8 (m), 1359.8 (m), 1222.8 (w), 1170.7 (w), 1082 (s), 781.1 (m), 680.3 (m), 572.8 (m), 430.2 (w).

**X-ray Crystallography**. A suitable crystal was affixed to the end of a glass fiber using silicone grease and rapidly transferred to the goniostat where it was cooled for characterization and data collection. X-ray crystallographic data were collected on a Picker four-circle diffractometer at –160 °C. Details of the diffractometry, low-temperature facilities, and computational procedures employed by the Molecular Structure Center are available elsewhere.[72] The structure was solved by direct methods (SHELXTL) and Fourier methods, and refined on *F* by full-matrix least-squares cycles.

For complex **2·4H$_2$O**, a systematic search of a limited hemisphere of reciprocal space revealed monoclinic symmetry and systematic absences corresponding to the unique space group *P2$_1$/c*. After data collection (+*h*, ±*k*, ±*l*; 6° < 2θ < 50°) and processing, and averaging of reflections measured more than once ($R_{av}$ = 0.198), the structure was solved without serious problems. The Mn$_4$ cation lies at a center of inversion. In the final refinement cycles, non-hydrogen atoms were refined anisotropically.  Hydrogen atoms were included for the cation in fixed, idealized positions with thermal parameters of 1.0 plus the isotropic thermal parameter of the parent atom.  Final $R$ and $R_W$ values are listed in Table 1.

For complex **3·4H$_2$O**, a systematic search of a limited hemisphere of reciprocal space revealed monoclinic symmetry and systematic absences corresponding to one of the monoclinic space groups, *C2/c* or *Cc*.  Subsequent solution and refinement of the structure confirmed centrosymmetric *C2/c* to be the correct space group.  All non-hydrogen atoms were readily located and refined anisotropically; hydrogen atoms were included in calculated positions as fixed atom contributors in the final refinement cycles.  Final $R$ and $R_W$ values are listed in Table 1.

**Physical Methods**.  IR measurements were made on samples pressed into KBr pellets using a Nicolet Magna-IR 550 spectrophotometer.  DC magnetic susceptibility experiments were performed on powdered microcrystalline samples (restrained in eicosane to prevent torquing at high fields) with a Quantum Design MPMS5 SQUID magnetometer equipped with a 5.5 T (55 kG) magnet and capable of operating in the 1.7 - 400 K range.  Diamagnetic contributions to the susceptibility were corrected using Pascal's constants.  AC magnetic susceptibility experiments above 1.7 K were carried out on a Quantum Design MPMS2 SQUID magnetometer.  AC field strength can be varied from 0.001 to 5 G at frequencies ranging from 0.0005 to 1512 Hz.  The temperature range available is 1.7 to 400 K.  Low temperature AC magnetic susceptibility experiments down to millikelvin temperature range were performed at the Institute of Solid State Physics, The University of Tokyo.  Details of the measurement and the experimental set up are described elsewhere.[69b]

High-frequency and –field EPR spectra were recorded at the National High Magnetic Field Laboratory, Tallahassee, Florida. The spectrometer is described in detail elsewhere[73] and is equipped with a 15/17 T superconducting magnet and can achieve temperatures ranging from 4 to 300 K. In this work, a Gunn diode, operating at a fundamental frequency of 95 ± 3 GHz, was used. The operating frequency range is 95 to 490 GHz. Frequencies between *ca*. 180 and 490 GHz were obtained by using a solid-state harmonic generator that multiplies the fundamental frequency (*i.e.*, 95 GHz) and high pass filters to filter out the lower frequency harmonics, however, the higher frequency harmonics pass through. The magnetic field was modulated at a typical frequency of 8.5 kHz and amplitude of 1.5 mT for the purpose of phase-sensitive detection. The magnetic field was swept at a rate of 0.2 T/min over a 0 – 14.5 T range.

**Results and Discussion**

**Synthesis**. Nitrogen-oxygen containing chelates such as picolinic acid, 2-hydroxymethyl-pyridine, 2-hydroxyethyl-pyridine, and pyridine-2,6-dimethanol are known to stabilize polynuclear manganese complexes with nuclearities of four,[69,74] seven,[75] and eighteen.[76] Recently, we reported a new structural type of a polynuclear manganese complex $[Mn_4(pdmH)_6(O_2CCH_3)_2](ClO_4)_2$ **(1)** which was obtained in the reaction of pyridine-2,6-dimethanol (pdmH) with a $\mu_3$-oxo-bridged trinuclear manganese carboxylate complex.[69] The complex was structurally determined to be tetranuclear with 2 $Mn^{II}$ and 2 $Mn^{III}$ ions. Unlike previously characterized tetranuclear Mn complexes,[74] the bridging is derived from the oxygen atom of the ligand; no bridging oxides are present as found in most polynuclear Mn complexes.

The magnetic properties of this complex were fully characterized employing the DC and AC magnetic susceptibility and HFEPR techniques. It was found to be a new example of a SMM.

It was reasoned that a family of this structural type could be found through the use of similar nitrogen-oxygen containing chelates such as 2-hydroxymethyl-pyridine (hmpH) and 6-methyl,2-hydroxymethyl-pyridine (6me-hmpH), ligands with only one methanolic substituent on the pyridine ring in contrast to the two on the pdmH$^-$ ligand. Initially, parallel reactions using hmpH instead of pdmH$_2$ with the trinuclear manganese carboxylate source produced a decanuclear manganese complex.[77] Thus, other routes were tried and it was found that reactions using simple mononuclear manganese complexes such as MnCl$_2$ or MnBr$_2$ in the presence of hmpH produced discrete tetranuclear complexes. It was concluded that the two new complexes resemble complex **1** based on initial IR data, but the structures were confirmed by X-ray crystallography.

**Descriptions of Structures**. Labeled ORTEP plots of the cation of [Mn$_4$(hmp)$_6$Br$_2$(H$_2$O)$_2$]Br$_2$·4H$_2$O **(2)** and [Mn$_4$(6me-hmp)$_6$Cl$_4$]·4H$_2$O **(3)** are given in Figures 3 and 4, respectively, and selected interatomic distances and angles are listed in Tables 2 and 3, respectively. Complex **2**·4H$_2$O crystallizes in the monoclinic space group *P*2$_1$/c with the cation lying on an inversion center and consisting of a planar Mn$_4$ rhombus. The unit cell contains 2 Br$^-$ counterions and 4 solvate water molecules per Mn$_4$ unit. Complex **2** is mixed-valence consisting of 2 Mn$^{II}$ and 2 Mn$^{III}$ ions and is similar to the tetranuclear manganese complex containing the ligand pdmH$^-$, complex **1**. Unlike **1** which has acetate as bridging ligand, the bidentate 2-hydroxymethylpyridine (hmp) ligand serves as the only bridging ligand for complex **2**. The Mn$_4$ rhombus in complex **2** is composed of two Mn$_3$ triangular faces held together by a $\mu_3$-oxygen atom [O(21)] of hmp$^-$. The edges of each triangular face consists of a $\mu_2$-oxygen [O(13)] and [O(5)]. Mn(1) and Mn(2) are assigned as the Mn$^{II}$ and Mn$^{III}$ ions, respectively, on the basis of bond valence sum calculations and the presence at Mn(2) of a Jahn-Teller elongation axis [O(21)-Mn(2)-N(12)].

The analysis of the X-ray data shows closely positioned Mn$_4$ clusters in the unit cell. Figure 5 shows the view of the packing of the Mn$_4$ molecule of complex **2** in the unit cell. The closest internuclear distance between the coordinated Br anions of the Mn$_4$ cluster is 4.91 Å. In addition to coordinated Br atoms, there are Br counterions very near the Mn$_4$ molecule. The closest separation between a Br counterion and the coordinated Br of an Mn$_4$ is 5.25 Å. There are also 4 closely positioned water molecules per Mn$_4$ molecule in the unit cell. The distance between a water molecule and a coordinated Br is 2.76 Å. Water molecules are also found to be closely positioned to a coordinated water of an Mn$_4$ molecule. The distance between a solvate water and a coordinated water is 3.49 Å. The 2.76 Å and 3.49 Å separations are well within the range for hydrogen bonding. Thus, from the manner in which the Mn$_4$ molecules are packed crystallographically, intermolecular interactions between neighboring Mn$_4$ molecules can be appreciable.

Careful analysis of the X-ray data also reveals that the Jahn-Teller axis [O(21)-Mn(2)-N(12)] is slightly bent. The angle formed by the O(21)-Mn(2) and the Mn(2)-N(12) bonds is 158° and not 180° expected if the axis is perfectly straight. In addition, two of the Mn$_4$ molecules in the unit cell are mutually related by a glide plane. The result is that the easy axes of two of the Mn$_4$ clusters in the unit cell of complex **2** are not exactly parallel (see Figure 6). The angle between the two molecular easy axes is close to 9.0 ± 0.5°. This slight canting of the easy molecular axis is important to consider in the analysis of HFEPR data.

Complex **3**·4H$_2$O crystallizes in a monoclinic space group *C*2/*c* and is best described as having four manganese ions arranged in a bent chain. The complex is also mixed-valence containing 2 Mn$^{II}$ ions and 2 Mn$^{III}$ ions but has no counterions as the charge is balanced by the coordinated ligands (Figure 4). Bridging between the metal centers is accomplished by the μ$_2$-oxygen of the 6-methyl-2-hydroxymethylpyridine anionic ligand; the chloride ions are terminal and are only bonded to the outer Mn ions of the chain. The central two Mn ions, Mn(1), are identified as being Mn$^{III}$ based on bond valence sum calculations and the presence of a Jahn-Teller elongation axis [O(3)-Mn(1)-N(10)].

The 3-dimensional structures shows 4 solvate molecules per molecule of complex **3** (Figure 7). The closest contact made between a solvate water molecule and a Mn$_4$ unit is 3.2 Å. The Mn$_4$ unit of complex **3** is tightly packed in row-like fashion. There are close Van der Waal contacts (3.5 – 4.2 Å) between a pyridine ring of the 6me-hmp ligand of neighboring Mn$_4$ units. Thus, it is possible that there are intermolecular interactions between molecules of complex **3**.

**DC Magnetic Susceptibility**. Variable temperature magnetic susceptibility measurements were carried out on microcrystalline powder samples of **2** and **3** restrained in eicosane. The data were collected under a 10 kG applied magnetic field in the 5-300 K range. Figure 8 shows the data plotted as $\chi_M T$ *vs*. temperature for complex **2**. The magnetic properties of **2** are similar to those observed for ferromagnetically coupled alkoxide bridged complexes[77] and the hydrated form of complex **1**.[69b] In fact, the $\chi_M T$ *vs*. *T* plot of complex **2** was found to be virtually superimposable to that of **1·2.5H₂O**. As shown in Figure 8, the value of $\chi_M T$ increases from 15.5 cm³ mol⁻¹ K (spin only ~ 14.75 cm³ mol⁻¹ K) at room temperature to a maximum of 33.6 cm³ mol⁻¹ K at 11 K and then decreases slightly to 25.9 cm³ mol⁻¹ K at 5 K. Ferromagnetic interactions are the dominant interactions as indicated by the increase in the $\chi_M T$ with decreasing temperature and are the origin of the high spin ground state of complex **2**. The slight decrease in $\chi_M T$ at lowest temperatures is likely the result of zero-field splitting in the ground state.

The $\chi_M T$ *vs.* temperature data for complex **2** were fit in order to determine the magnitudes of the ferromagnetic exchange interactions. It was expected that the ferromagnetic interactions in complex **2** would be nearly identical to that found for complex **1·2.5H₂O**. The same procedure used to fit the data for **1·2.5H₂O** was used to fit the data shown in Figure 8. Because of the near identical core structure between complex **2** and **1·2.5H₂O**, a model which considers only two dominant exchange pathways was used as shown in Figure 9. This approximation equates $J_{1'2} = J_{1'2'} = J_{12} = J_{12'} = J_{wb}$ and the Kambe vector coupling method can be used to solve the spin Hamiltonian which is given in equation (1):

$$\hat{H} = -J_{wb}(\hat{S}_T^2 - \hat{S}_A^2 - \hat{S}_B^2) - J_{bb}(\hat{S}_A^2 - \hat{S}_1^2 - \hat{S}_3^2) \qquad (1)$$

where: $\hat{S}_A = \hat{S}_1 + \hat{S}_3$, $\hat{S}_B = \hat{S}_2 + \hat{S}_4$, and $\hat{S}_T = \hat{S}_A + \hat{S}_B$.

The Kambe method gives the eigenvalue expression in equation (2):

$$E(S_T) = -J_{wb}[S_T(S_T+1) - S_A(S_A+1) - S_B(S_B+1)] \\ - J_{bb}[S_A(S_A+1)] \quad (2)$$

With two $s = 2$ [Mn(III)] and two $s = 5/2$ [Mn(II)] interacting ions present in complex **2** there are a total of 110 possible states with $S_T$, the total spin of the Mn$_4$ cluster, ranging from 0 to 9.

This eigenvalue expression and the corresponding Van Vleck equation were then used to least-squares fit the experimental data measured in the 5-300 K range (Figure 8). Data below 15 K were omitted in the fitting because magnetization saturation effects likely dominate in this temperature range. With the temperature independent paramagnetism (TIP) fixed at 600 x 10$^{-6}$ cm$^3$ mol$^{-1}$, a good fit was achieved (solid line) with final optimized parameters being: $g = 1.94$, $J_{wb}/hc = 0.936$ cm$^{-1}$, and $J_{bb}/hc = 8.848$ cm$^{-1}$. These values correspond to a ground state with $(S_T, S_A, S_B) = (9, 4, 5)$. It must be emphasized that the magnitudes of the exchange interactions in complex **2** are comparable to those found for complex **1·2.5H$_2$O** resulting in a spin ground state of $S = 9$.[69b]

The variable temperature susceptibility data for complex **3** also show signs of ferromagnetic interactions. Figure 10 gives the data plotted as $\chi_M T$ vs. temperature. The value of $\chi_M T$ increases from 16.3 cm$^3$ mol$^{-1}$ K at 300 K to around 22.6 cm$^3$ mol$^{-1}$ K at 30 K before decreasing to 16.4 cm$^3$ mol$^{-1}$ K at 5 K. Compared to the data for complex **2**, a broader maximum is observed in the plot of $\chi_M T$ vs. temperature for complex **3** at low temperatures. However, the similar increase in $\chi_M T$ at low temperatures suggests that the ground spin state for **3** is also large. Zero-field splitting is also present in **3** as evident in the sharp decrease in $\chi_M T$ at low temperatures.

It can be seen in Figure 11 that there are only two different exchange pathways in complex **3**. The two terminal $Mn^{II}$-$Mn^{III}$ interactions are equivalent from symmetry considerations. Thus, it is possible to use the Kambe model to fit the $\chi_M T$ vs. temperature shown in Figure 10. With the temperature independent paramagnetism (TIP) fixed at 600 x 10$^{-6}$ cm$^3$ mol$^{-1}$, a good fit was achieved (solid line) with final optimized parameters being: $g = 1.95$, $J_{wb}/hc = 0.593$ cm$^{-1}$, and $J_{bb}/hc = 16.17$ cm$^{-1}$. These values correspond to a ground state with $(S_T, S_A, S_B) = (9, 4, 5)$. The $J_{wb}$ interactions are nearly twice as small compared to that for complex **2**, but the $J_{bb}$ interactions are nearly twice as large.

**Variable Field DC Magnetization Measurements**. The large spin ground states and zero-field splittings of complexes **2** and **3** were further examined through variable field magnetization measurements. Saturation of magnetization is achieved at high fields and low temperatures; the field dependence of the magnetization can be analyzed to obtain the spin of ground state and the magnitude of the zero-field splitting. For complexes with no zero-field splitting, the plot of magnetization versus external magnetic field follows a simple Brillouin function where the magnetization saturates at a value of $gS$ and the isofield lines superimposed. Magnetization measurements were carried out for complexes **2** and **3** in 20 - 50 kG fields and in the 2.0 - 4.0 K range. The results for complex **2** (top) and complex **3** (bottom) are plotted as reduced magnetization $M/N\mu_B$ vs. $H/T$ in Figure 12, where $M$ is the magnetization, $N$ is Avogadro's number, $\mu_B$ is the Bohr magneton, and $H$ is the magnetic field. It is apparent that complexes **2** and **3** both have ground states with a large spin and appreciable zero-field splitting.

At the highest field and lowest temperature complexes **2** and **3** have saturation of magnetization values of 16.40 $N\mu_B$ and 16.45 $N\mu_B$, respectively. Zero-field splitting is evident by the non-superposition of the isofield lines. A saturation of magnetization value of 16.4 $N\mu_B$ is similar to that observed for complex **1**·2.5H$_2$O, which was determined to have a ground state spin of $S = 9$. One notable difference between the data for complexes **2** and **3** is seen. The isofield lines for complex **2** are essentially flat whereas those for complex **3** show some curvature. This suggests that in the field and temperature ranges of the experiment, saturation of magnetization for complex **2** is achieved, whereas it is not complete for complex **3**. This may be due to population of excited states with smaller spin than the ground state in complex **3**.

To obtain the ground state and $D$ values from the variable field magnetization data, the data sets were fit using a model which assumes that the ground state is well-isolated and is the only state populated in the 2-4 K and 20-50 kG ranges. A spin Hamiltonian including the Zeeman interaction and axial zero-field spitting ($D\hat{S}_z^2$) was used to least-squares fit the data. The matrix was diagonalized on each cycle and a powder average was calculated.[79] The results of the fitting are shown as solid lines in Figure 12. The fitting of the data assuming a $S = 9$ ground state produced nearly the same $g$ and $D$ parameters as were obtained for **1**·2.5H$_2$O.[69b] The fit of the data for complex **2** gave $g = 1.95$ and $D/k_B = -0.34$ K as the final optimized parameters for a spin ground state of $S = 9$. For complex **3**, the fit gave $g = 1.95$ and $D/k_B = -0.25$ K as the final optimized parameters for a ground state of $S = 9$. Although the $g$ and $D$ values obtained for complex **3** are reasonable, it is clear from inspection that the fit of the data for complex **3** is not as good as obtained for complex **2**. This lends support to the suggestion that population of small spin excited states as the origin for the incomplete saturation observed for complex **3**.

**High-frequency EPR Spectroscopy**. High-frequency EPR (HFEPR) measurements were made to further characterize the ground states of complexes **2** and **3**. In addition to the importance of obtaining direct measurements of both $S$ and $D$ values, it was of interest to compare the HFEPR spectra for complexes **2** and **1**·2.5H$_2$O. HFEPR is useful in the determination of both the $S$ and $D$ values of polynuclear complexes with large spin and magnetic anisotropy.[80] Transitions between the $M_S$ levels arising from zero-field splitting in the ground state (see Figure 13) can be observed. These fine structure transitions can be analyzed to obtain both the $S$ and $D$ values directly; the sign of the zero-field splitting parameter $D$ can also be determined. The $S$ value can be qualitatively determined by counting the number of the resonances if the fine structure is well resolved and there is enough thermal energy to populate all levels. For a complex with a given $S$ value for instance, a total of $2S$ resonances will be observed. The $D$ value on the other hand is obtained from the spacing between the resonances and the sign of $D$ is obtained from the temperature dependence of the intensities of the fine structure peaks. Thus, a complex with a negative $D$ value will give an EPR spectrum with the number of parallel transitions increasing towards high fields when the temperature is increased. The trend is reversed for a complex with a positive $D$ value.

The spin Hamiltonian that describes a molecule with a spin $S$ influenced by Zeeman interaction and axial zero-field splitting is expressed in equation (3):

$$\hat{H} = g\mu_B \hat{S} \cdot H + D[\hat{S}_z^2 - (1/3)S(S+1)] \qquad (3)$$

where $g$ is the Lande factor and $\mu_B$ is the Bohr magneton. In this equation, the first term is the Zeeman term and the second term gauges the axial zero-field splitting in the ground state. The resonance lines in a HFEPR spectrum for large spin molecules with uniaxial anisotropy depend on the orientation of the crystallites in the presence of the high magnetic field. The position of these resonances is given by:

$$H_r = h\nu/g\mu_B - (2M_s+1)D' \qquad (4)$$

where $h\nu$ is the microwave energy and $D' = [3(\cos^2\theta) - 1] \cdot D/2g\mu_B$ from first-order approximation with $g$ and $\mu_B$ as the usual physical constants and $\theta$ being the angle between the easy axis of the molecule in the crystal and the magnetic field. Since single-molecule magnets have such a large magnetic anisotropy, the crystallites usually torque in the presence of strong magnetic fields such that the anisotropy axes of the molecules in the crystal align parallel to the magnetic field. $\theta$ is zero if all the anisotropy axes are parallel to the field, however, depending on the orientation of the molecules relative to one another within the crystal, there can be an angle between the molecular anisotropy axis and the field. The angle, if large enough, can alter the positions of the parallel resonance in the HFEPR spectrum and influence the magnitude of $D$.

    HFEPR measurements using the frequencies of 191.30, 291.26, and 388.35 GHz were made for a loosely packed microcrystalline powder sample of complex **2** to allow torquing. Shown in Figure 14 are quasi-single crystal spectra at 291.26 GHz and either 30 or 60 K. In the 30 K spectrum, two sets of resonances are observed: one set occurring in the 5-8 T region and other occurring in the 9-12 T region. Most of the resonances in the 9-12 T region are due to the

higher frequency harmonics (i.e., 4 times the 95.65 GHz). Independent measurements at 388.35 GHz confirmed that the resonances in the 9-12 T region of the 291 GHz spectrum are from the higher harmonics. In the 30 K spectrum, admixture of absorption and derivative signals due to rapid passage effects characteristic of one-path transmission method complicates the appearance of the spectrum and some of the resonances are not clearly defined. However, it is still evident that there are regularly spaced resonances. These resonances become more evident and well defined at 60 K. At 60 K, six well-defined resonances are observed in the 5-8 T region as denoted by an asterisk. There is appreciable thermal energy at 60 K to populate higher $M_S$ levels, thus more resonances are seen at this temperature. Comparing the spectra at the two temperatures, the resonances observed in the high field region in the 60 K spectrum have larger relative intensities compared to those in the 30 K spectrum. Thus, it can be concluded that $D < 0$ for complex **2**. If $D > 0$ for complex **2**, the reverse trend would be observed.

HFEPR measurements were also made for a loosely packed microcrystalline powder sample of complex **3** to see if fine structure resonances are observable for this complex. Shown in Figure 15 is a plot of the HFEPR spectra for complex **3** at two different frequencies 190.01 and 285.02 GHz at 60 K. The peaks in the 2-6 T region of the 190 GHz spectrum are attributed to fine structure transitions, however, the poor resolution limited a detailed study into the temperature dependence. One possible reason for the poor resolution is incomplete torquing of the sample to give well-resolved parallel signals. Even for magnetically anisotropic complexes such as complex **3**, the manner in which the sample is packed for measurement as well as the size of the crystallites used can impact the torquing in a magnetic field. Because parallel EPR spectra are observed in the field region shown, incomplete torquing of the sample can dramatically influence the lineshape and hence resolution of the transitions. Another possibility for the poor resolution is that the spectrum contains signals from chemical or physical impurities. The presence of a physical impurity is supported by the large $g = 2$ signal. Detailed experiments are necessary to better understand the origins of the lineshape and resolutions in the EPR spectrum.

The well-resolved fine structure in the spectra for complex **2** are more amenable for more detailed analysis. Shown in Figure 16 are the results of measurements made at three different frequencies at 60 K. Regularly spaced transitions attributed to fine structure observed at 191.3 GHz are also seen at 291.26 and 388.35 GHz. Table 3.4 summarizes the observed resonance positions for the HFEPR spectra for complex **2** at the three frequencies. These resonances correspond to transitions from the $M_S$ to the ($M_S$+1) levels of the ground state. Thus, assuming complex **2** has a ground state spin of $S = 9$ with zero-field splitting, the 1.938 T resonance in the 191.3 GHz spectrum corresponds to the $M_S = -9 \rightarrow M_S = -8$ transition, the 2.308 T resonance corresponds to a $M_S = -8 \rightarrow M_S = -7$ transition, and so on. In the 191.3 GHz spectrum at 60 K, all the transitions up to the $M_S = 0 \rightarrow M_S = +1$ are observed. The other 7 fine structure transitions from $M_S = +1 \rightarrow M_S = +2$ up to $M_S = +8 \rightarrow M_S = +9$ are not observed in part because at 60 K there is little Boltzmann population in these higher energy levels. It is also possible that the upper resonances are obscured by the presence of extra signals. A broad signal is observed at *ca*. 4.5 and 6.5 T in the 191.30 GHz spectrum (Figure 16). These signals also appear in the higher harmonics; they are more apparent in the 291.26 GHz spectrum than the 388.35 GHz spectrum. These signals may arise from defect sites in the crystal of complex **2** and are real. The presence of these signals can obscure the fine structure transitions at high fields. The less resonances seen at 388.35 GHz compared to 191.30 GHz is due to less microwave power available resulting in poor signal to noise ratio. Thus, fewer peaks are seen in the 388.35 GHz spectrum compared to the 191.30 GHz spectrum.

Upon close inspection of the resonance lines at each frequency in Figure 16, it was found that they are not equally spaced. For example, in the 191.3 GHz spectrum the spacing between the resonances increases toward the high field region of the fine structure pattern. This is further evident when the resonance field at which the transitions occur is plotted as a function of $M_S$. Shown in Figure 17 is the plot of resonance field *vs.* $M_S$ for the data measured at the three frequencies. This plot is made assuming that the first peak in the fine structure is due to a transition from $M_S = -9 \rightarrow M_S = -8$. It is clear from this plot that the resonance field positions do *not* exhibit a linear dependence on $M_S$, as expected from equation (4). As a result, inclusion of the longitudinal quartic zero-field splitting term, $(35B_4^0 \hat{S}_z^4)$, to equation (3) was needed to analyze the data.

$$\hat{H} = g\mu_B \hat{S} \cdot H + D[\hat{S}_Z^2 - (1/3)S(S+1)] + 35B_4^0 \hat{S}_Z^4 \quad (5)$$

The resonance field data for all three frequencies were analyzed using equation (5) by adopting the eigenfield approach with downhill simplex method.[81] The solid lines in Figure 17 correspond to the fit of the data. Because the easy axes of neighboring molecules of complex **2** are canted at $9.0 \pm 0.5°$, the angle between the easy axis of each molecule and the applied field is 4.5°. This angle was taken into account in the analysis. The fit was optimal when all the parameters, $g$, $D$, and $B_4^0$ were varied, which resulted in $g = 2.03$, $D/k_B = -0.497$ K and $B_4^0/k_B = 1.76 \times 10^{-5}$ K. However, these results are not reasonable, for the final $g$ value is greater than 2.00. It is typical to have $g \leq 2$ for manganese complexes.[82] A somewhat poorer fit was obtained by fixing the $g$ value at 1.99 which resulted in $D/k_B$ and $B_4^0/k_B$ values of -0.527 K and 2.18x10$^{-5}$ K respectively. At this point, it was decided that the HFEPR spectra needed to be simulated.

**Simulation of HFEPR Spectra.** The spectra shown in Figures 14 and 16 are parallel spectra arising from torquing of the sample in the magnetic field. Thus, it is expected that only the $D$ and $B_4^0$ parameters influence the spectra. Shown in Figure 18 is the plot of the experimental spectrum (top) together with a simulation (bottom) for the 191.30 GHz spectrum at 60 K. The canting angle of 4.5° was taken into account in the simulation. The following parameters were initially used to simulate the spectra for a complex with $S = 9$ ground state: $D/k_B$ = -0.498 K, $g$ = 1.999, and $B_4^0/k_B$ = 1.72x10$^{-5}$ K. These parameters are close to those obtained above by fitting the resonance positions to equation (5). The simulated spectrum is seen to well reproduce the experimental resonance positions. Figure 19 shows the simulations of the 291.26 and 388.35 GHz spectra for complex **2**. The simulations were made using the same parameters as those for the 191.3 GHz spectrum. As can be seen, one set of parameters gives good simulations for all the spectra obtained for complex **2**.

Although the simulation reproduced the experimental resonance positions accurately, there is some disagreement in the amplitudes of the individual resonance lines between experiment and simulations. The amplitude of the experimental resonance lines decreases much more rapidly toward high fields than seen in the simulation. This discrepancy may be due to resonance linewidth. The linewidth of the $M_S$ = -9 to $M_S$ = -8 transition is measured to be *ca.* 600 G broad. The $M_S$ = -3 to $M_S$ = -2 transition is *ca.* 900 G broad. The linewidth of the transitions observed above 10.0 T, assuming that they are a part of this series, is measured to be 2000 G and up to 3000 G broad. The simulation program does not account for the change in linewidth at high fields, but assumes a constant linewidth for the resonances. This is a reason for the reduced amplitude at high fields in the experimental spectrum. The origin of the broadening of the linewidth is not clear, but perhaps could be due to relaxation dependence on field. Additional experiments are required which are presently ongoing.

Because the HFEPR spectra recorded are parallel spectra, it was not possible to determine the rhombic zero-field splitting parameter $E$ with precision. $E$ mostly affects perpendicular transitions which can be seen in powder EPR spectra where both the parallel and perpendicular transitions are observed. It would be of interest to know the $E$ parameter for complex **2** as $E$ is believed to have a profound effect on the rate of magnetization tunneling in SMM's. The rhombic zero-field interaction, $E(\hat{S}_x^2 - \hat{S}_y^2)$, can mix together $M_S$ states that differ in 2 in $M_S$ values. Thus, this interaction mixes the $M_S = -9$ state with the $M_S = -7$ state, or in general a $M_S$ state and a ($M_S \pm 2$) state. Together with other transverse interaction terms such as transverse (perpendicular) magnetic field, rhombic zero-field splitting determine the rate of magnetization tunneling between the $2S + 1$ levels in the ground state (Figure 1). Analysis of powder EPR spectra or inelastic neutron scattering measurements should give the precise value of $E$ for complex **2**.

**AC Magnetic Susceptibility**. Variable-field magnetization data and HFEPR spectra showed that complex **2** has a $S = 9$ ground state. Since the $S = 9$ ground state has considerable magnetic anisotropy with $D/k_B = -0.498$ K, a potential energy barrier is present for the reversal of the direction of magnetization in complex **2**. As illustrated in Figure 1, the presence of $D$ splits the $S = 9$ ground state into $M_S = \pm 9, \pm 8,....., \pm 1, 0$ levels. When $D < 0$, in zero applied field the $M_S = +9$ and $M_S = -9$ levels are the lowest in energy and $M_S = 0$ the highest. Complex **2** should exhibit slow magnetization relaxation in response to a change in direction of the external magnetic field.

AC susceptibility measures the magnetization under an oscillating magnetic field, so it is possible to determine the rate at which a single-molecule magnet reverses its direction of magnetization. A collapse in the in-phase ($\chi_M'$) component and an appearance of a peak in the out-of-phase ($\chi_M''$) component of the AC magnetic susceptibility are observed when the frequency of the oscillating magnetic field equals the rate at which a single-molecule reverses its direction of magnetization. The initial observations that complex **2** functions as a single-molecule magnet were made with AC susceptibility measurements in the 1.8 - 6.4 K range employing an MPMS2 magnetometer. These revealed the onset of an out-of-phase AC signal as shown in Figure 20 where plots of both the in-phase ($\chi_M'$) and the out-of-phase ($\chi_M''$) AC signals are given. It can be seen that at the lowest temperatures a frequency dependent $\chi_M''$ signal is seen.

These initial observations were followed by AC magnetic susceptibility measurements made on a polycrystalline sample of complex **2** in the 0.05 - 4.0 K range. Figure 21 gives the results of these AC susceptibility measurements at three different frequencies. At 992 Hz, the in-phase component is about 45 cm$^3$ mol$^{-1}$ K at 3.5 K and it steadily decreases to about 30 cm$^3$ mol$^{-1}$ K at around 2 K. This decrease in the 2.0 – 3.5 K range is attributed to intermolecular antiferromagnetic interactions the effects of which can be observed at low temperatures. This supports what is observed from the X-ray structure for complex **2** which reveal the presence of hydrogen bonding contacts present within the crystal and also small separation between molecules of complex **2**. Such a large drop in the in-phase component was not observed in the analogous complex **1**;[69b] the in-phase component of the AC magnetic susceptibility remained essentially flat at high temperature prior to the onset of spin freezing. Below 2 K, an abrupt drop in the in-phase component for complex **2** is observed from 30 cm$^3$ mol$^{-1}$ K to near 0 cm$^3$ mol$^{-1}$ K. The sudden drop in the in-phase component is attributed to a relaxation process as this is accompanied by a peak in the out-of-phase component. At 992 Hz, the peak in the out-of-phase component was observed at 1.5 K. This means that at 1.5 K, the rate at which the magnetization reverses its direction is equal to the frequency of the oscillating AC magnetic field. Measurements were made at six different frequencies in the 10.5 - 992 Hz range to determine the rates of magnetization reversal at different temperatures. The results are summarized in Table 3.5. It must be noted that complex **2** has a faster rate of spin reversal than complex **1**·2.5H$_2$O, for the out-of-phase AC response for complex **2** occurs at a lower temperature than was observed for complex **1**·2.5H$_2$O.[69b] It is possible that the presence of intermolecular interaction is the reason for a faster rate of spin reversal in complex **2** compared to complex **1**·2.5H$_2$O. Intermolecular exchange interactions may serve as an off-diagonal matrix element responsible to quantum tunneling. It has a similar effect as dipolar interactions, which result from a transverse field exerted from neighboring molecules. Dipolar interactions have been proposed to play a big role in quantum tunneling.[83]

The magnetization relaxation data for complex **2** were analyzed using the Arrhenius law $\tau = \tau_0 e^{-\Delta E/k_B T}$. In this equation for the relaxation time $\tau$, $\Delta E$ is the activation energy, $k_B$ is Boltzmann constant and $\tau_0$ is the preexponential factor. Figure 22 gives a plot of ln(1/$\tau$) *vs.* 1/*T*, where the solid line represents the least-squares fit of the data to the Arrhenius equation. From this fit the activation energy $\Delta E$ was found to be 15.8 K with a preexponential factor of 2.93 x $10^{-8}$ s. This value of $\Delta E$ is to be compared with the thermodynamic barrier of $U/k_B = |D\hat{S}^2|/k_B$ = 40.3 K, calculated for complex **2** with the values of $S = 9$ and $D/k_B = -0.498$ K obtained from the simulation of the HFEPR spectra. It is reasonable that $U$ is greater than $\Delta E$, for the reversal of the direction of magnetization in an SMM may not involve only thermal activation over the potential-energy barrier, but also quantum tunneling of the direction of magnetization.

AC susceptibility measurements were also carried out for complex **3** in the 1.8 - 6.0 K range employing the MPMS2 magnetometer. Figure 23 shows the results of both the in-phase (top) and out-of-phase (bottom). At 1000 Hz, the in-phase component is about 21 $cm^3$ $mol^{-1}$ K at 3.5 K and it steadily decreases to about 12 $cm^3$ $mol^{-1}$ K at around 2 K. This decrease in the 2.0 – 3.5 K range is attributed to intermolecular antiferromagnetic interactions made possible via the Van der Waal interactions between the ligands of neighboring molecules of complex **3** in the crystal. Interestingly, no signs of a drop in the in-phase component and a rise in the out-of-phase component of the AC susceptibility were observed at 1.8 K. It is possible that the $D$ value for complex **3** is smaller than that in complex **2**. AC susceptibility measurements in the millikelvin are still in progress to determine whether there is evidence of slow magnetization relaxation at lower temperatures.

It is important to compare the values of $U$ and $\Delta E$ obtained for complex **2** with those obtained for other manganese SMM's. The analogous complex $[Mn_4(pdmH)_6(O_2CCH_3)_2](ClO_4)_2 \cdot 2.5H_2O$ (**1**·2.5H$_2$O) was determined[69b] with HFEPR data to have $S = 9$ ground state with $D/k_B = -0.45$ K. This gives a thermodynamic barrier of $U/k_B$ = 36.5 K for **1**·2.5H$_2$O compared to $U/k_B$ = 40.3 K for **2**. Similarly, **1**·2.5H$_2$O was reported to have an activation energy for reversal of magnetization of $\Delta E/k_B$ = 16.7 K, compared to a value of $\Delta E/k_B$ = 15.8 K for complex **2**. The smaller $\Delta E$ value for **2** is consistent with the observation of $\chi_M''$ signals at lower temperatures than for complex **1·2.5H$_2$O**, implying that **2** has a faster rate of magnetization reversal. The desolvated form of complex **1** has a $S = 8$ ground state with $D/k_B$ = -0.35 K and $U/k_B$ = 22.4 K and it has been determined with AC susceptibility data to have $\Delta E/k_B$ = 17.3 K. Preliminary magnetization relaxation measurements[70] show that the desolvated complex **1** exhibits a temperature-independent magnetization relaxation rate of 1 x 10$^{-4}$ s$^{-1}$. Detailed magnetization decay measurements are being carried out on complex **2**.

**Concluding Comments**

One of the goals of this research was to synthesize new manganese based molecules using nitrogen-oxygen chelate ligands analogous to pyridine-dimethanol (pdmH). The motivation behind the use of similar ligands stems from the discovery that a new manganese structural type was found with pdmH, complex **1**, and that this new complex was a single-molecule magnet. Two new types of manganese complexes are reported in this study. Complex **2** was obtained using hydroxymethyl pyridine (hmp) as the ligand and complex **3** was obtained using a methylated version of hmp, 6-methyl-hydroxymethyl pyridine. Although hmp and 6me-hmp are both structural analogues of pdmH, they give complexes with completely different core topologies. The cation in complex **2** is a tetranuclear manganese complex that bears close structural resemblance to the cation in complex **1**. However, complex **3** bears no resemblance to **1** and **2**; it has four Mn ions arranged in a bent chain.

The magnetization characteristics of both complexes **2** and **3** are reported. Both complexes have a $S = 9$ ground state. HFEPR measurements were made and the data for complex **2** were analyzed which showed that the $S = 9$ and $D/k_B = -0.498$ K values are comparable to those reported for **1·H₂O**. Complex **2** shows frequency-dependent out-of-phase AC susceptibility signals characteristic of single-molecule magnets. Interestingly, the maximum in the out-of-phase magnetic susceptibility signal observed with a 1 kHz oscillating field occurs at a lower temperature for complex **2** than for complex **1** and its hydrate.

Even though complex **2** and **1·H₂O** have comparable potential-energy barriers ($U/k_B = 36.5$ K for **1·H₂O** and $U/k_B = 40.3$ K for **2**), complex **2** exhibits a faster magnetization relaxation rate. This may be the result of rhombic zero-field splitting, $E(\hat{S}_x^2 - \hat{S}_y^2)$, for complex **2**. It may also be from transverse components of dipolar fields within the crystal of complex **2** as well as hyperfine interactions derived from nuclear spins in the molecule.

**Acknowledgement**. DNH and GC thank the National Science Foundation for support of this research.

**Supporting Information Available**. X-ray crystallographic files in CIF format. This material is available free of charge via the internet at http://pubs.acs.org.

**Table 1.** Crystallographic Data for [Mn$_4$(hmp)$_6$Br$_2$(H$_2$O)$_2$]Br$_2$·4H$_2$O **(2)** and [Mn$_4$(6me-hmp)$_6$Cl$_4$]·4H$_2$O **(3)**.

|  | (2) | (3) |
|---|---|---|
| Formula | C$_{36}$H$_{48}$Br$_4$Mn$_4$N$_6$O$_{12}$ | C$_{42}$H$_{56}$Cl$_4$Mn$_4$N$_6$O$_{10}$ |
| $a$, Å | 10.907(0) | 17.0852(4) |
| $b$, Å | 15.788(0) | 20.8781(5) |
| $c$, Å | 13.941(0) | 14.1835(3) |
| $\alpha$, ° | 90 | 90 |
| $\beta$, ° | 101.21(0) | 90.5485(8) |
| $\gamma$, ° | 90 | 90 |
| $V$, Å$^3$ | 2354.94 | 5059.1 |
| $Z$ | 2 | 4 |
| formula weight | 1296.18 g mol$^{-1}$ | 1166.50 g mol$^{-1}$ |
| space group | $P2_1/c$ | $C2/c$ |
| $T$, °C | -160 | -160 |
| $\lambda$, Å | 0.71069 [a] | 0.71073 [a] |
| $\rho_{cald}$, g/cm$^{-3}$ | 1.828 | 1.532 |
| $\mu$(MoK$_a$), cm$^{-1}$ | 45.142 | 12.450 |
| $R$ [b] | 0.030 | 0.038 |
| $R_w$ [c] | 0.025 | 0.043 |

[a] Graphite monochromator. [b] $R = \Sigma[\ |F_o| - |F_c|\ ]/\Sigma|F_o|$  [c] $R_w = [\Sigma w(\ |F_o| - |F_c|)^2/\Sigma w|F_o|^2]^{1/2}$ where $w = 1/\sigma^2(|F_o|)$.

**Table 2.** Selected Interatomic Distances (Å) and Angles (°) for [Mn$_4$(hmp)$_6$Br$_2$(H$_2$O)$_2$]Br$_2$·4H$_2$O.

___

| | | | | | | | |
|---|---|---|---|---|---|---|---|
| Mn(1) | O(4) | | 2.226(2) | Mn(2) | O(5) | | 1.8882(17) |
| Mn(1) | O(5) | | 2.1950(18) | Mn(2) | O(13) | | 1.8770(17) |
| Mn(1) | O(13) | | 2.1477(18) | Mn(2) | O(21) | | 1.9545(18) |
| Mn(1) | O(21) | | 2.2952(17) | Mn(2) | O(21) | | 2.2563(18) |
| Mn(1) | N(28) | | 2.235(2) | Mn(2) | N(12) | | 2.217(2) |
| Mn(1) | Br(3) | | 2.554 | Mn(2) | N(20) | | 2.048(2) |
| | | | | | | | |
| O(4) | Mn(1) | O(5) | 168.89(8) | O(5) | Mn(2) | N(20) | 97.54(8) |
| O(4) | Mn(1) | O(13) | 83.66(9) | O(13) | Mn(2) | O(21) | 83.51(8) |
| O(4) | Mn(1) | O(21) | 95.71(8) | O(13) | Mn(2) | O(21) | 95.75(7) |
| O(4) | Mn(1) | N(28) | 83.33(9) | O(13) | Mn(2) | N(12) | 106.07(8) |
| O(4) | Mn(1) | Br(3) | 94.61(0) | O(13) | Mn(2) | N(20) | 80.66(8) |
| O(5) | Mn(1) | O(13) | 92.02(7) | O(21) | Mn(2) | O(21) | 81.65(7) |
| O(5) | Mn(1) | O(21) | 73.18(6) | O(21) | Mn(2) | N(12) | 97.67(8) |
| O(5) | Mn(1) | N(28) | 93.48(8) | O(21) | Mn(2) | N(12) | 157.99(7) |
| O(5) | Mn(1) | Br(3) | 96.49(0) | O(21) | Mn(2) | N(20) | 163.53(8) |
| O(13) | Mn(1) | O(21) | 70.03(6) | O(21) | Mn(2) | N(20) | 95.39(8) |
| O(13) | Mn(1) | N(28) | 139.34(8) | N(12) | Mn(2) | N(20) | 91.08(9) |
| O(13) | Mn(1) | Br(3) | 113.39(0) | Mn(1) | O(5) | Mn(2) | 119.19(8) |
| O(21) | Mn(1) | N(28) | 73.18(7) | Mn(1) | O(13) | Mn(2) | 107.24(8) |
| O(5) | Mn(2) | O(13) | 175.21(8) | Mn(1) | O(21) | Mn(2) | 95.66(6) |
| O(5) | Mn(2) | O(21) | 97.88(8) | Mn(1) | O(21) | Mn(2) | 99.22(8) |
| O(5) | Mn(2) | O(21) | 79.96(7) | Mn(1) | O(21) | Mn(2) | 98.35(7) |
| O(5) | Mn(2) | N(12) | 78.34(8) | | | | |

___

**Table 3.** Selected Interatomic Distances (Å) and Angles (°) for [Mn$_4$(6me-hmp)$_6$Cl$_4$]·4H$_2$O.

| | | | | | | | |
|---|---|---|---|---|---|---|---|
| Mn(1) | O(3) | | 1.909(2) | Mn(2) | Cl(30) | | 2.3782(9) |
| Mn(1) | O(3) | | 2.194(2) | Mn(2) | Cl(31) | | 2.3768(10) |
| Mn(1) | O(12) | | 1.919(2) | Mn(2) | O(12) | | 2.092(2) |
| Mn(1) | O(21) | | 1.867(2) | Mn(2) | O(21) | | 2.184(2) |
| Mn(1) | N(10) | | 2.264(3) | Mn(2) | N(19) | | 2.288(3) |
| Mn(1) | N(28) | | 2.146(3) | | | | |
| | | | | N(10) | Mn(1) | N(28) | 89.94(10) |
| O(3) | Mn(1) | O(3) | 74.18(9) | Cl(30) | Mn(2) | Cl(31) | 115.35(4) |
| O(3) | Mn(1) | O(12) | 104.32(9) | Cl(30) | Mn(2) | O(12) | 125.47(7) |
| O(3) | Mn(1) | O(12) | 94.34(9) | Cl(30) | Mn(2) | O(21) | 94.98(6) |
| O(3) | Mn(1) | O(21) | 98.95(9) | Cl(30) | Mn(2) | N(19) | 107.06(7) |
| O(3) | Mn(1) | O(21) | 168.54(10) | Cl(31) | Mn(2) | O(12) | 118.91(7) |
| O(3) | Mn(1) | N(10) | 148.54(9) | Cl(31) | Mn(2) | O(21) | 103.59(7) |
| O(3) | Mn(1) | N(10) | 77.49(9) | Cl(31) | Mn(2) | N(19) | 95.27(8) |
| O(3) | Mn(1) | N(28) | 86.02(9) | O(12) | Mn(2) | O(21) | 67.92(8) |
| O(3) | Mn(1) | N(28) | 107.93(10) | O(12) | Mn(2) | N(19) | 73.41(9) |
| O(12) | Mn(1) | O(21) | 78.28(9) | O(21) | Mn(2) | N(19) | 141.32(9) |
| O(12) | Mn(1) | N(10) | 91.09(9) | Mn(1) | O(3) | Mn(1) | 102.43(9) |
| O(12) | Mn(1) | N(28) | 157.38(10) | Mn(1) | O(12) | Mn(2) | 107.75(10) |
| O(21) | Mn(1) | N(10) | 111.09(10) | Mn(1) | O(21) | Mn(2) | 106.04(10) |
| O(21) | Mn(1) | N(28) | 80.28(10) | | | | |

**Table 4.** Resonant Field Transitions for $[Mn_4(hmp)_6Br_2(H_2O)_2]Br_2 \cdot 4H_2O$.

| Transition | Resonance Fields | | | | | |
|---|---|---|---|---|---|---|
| | 191.3 GHz | | 291.26 GHz | | 388.35 GHz | |
| $M_S \to M_S + 1$ | exp | calc | exp | calc | exp | calc |
| $-9 \to -8$ | 1.938 | 1.8796 | 5.437 | 5.4687 | 8.873 | 8.9544 |
| $-8 \to -7$ | 2.308 | 2.2073 | 5.797 | 5.7963 | 9.232 | 9.2821 |
| $-7 \to -6$ | 2.73 | 2.6321 | 6.222 | 6.2212 | 9.656 | 9.7069 |
| $-6 \to -5$ | 3.232 | 3.1413 | 6.708 | 6.7303 | 10.161 | 10.2161 |
| $-5 \to -4$ | 3.787 | 3.7217 | 7.275 | 7.3107 | 10.73 | 10.7965 |
| $-4 \to -3$ | 4.405 | 4.3604 | 7.913 | 7.9495 | | |
| $-3 \to -2$ | 5.069 | 5.0446 | 8.616 | 8.6336 | | |
| $-2 \to -1$ | 5.75 | 5.7611 | 9.205 | 9.3502 | | |
| $-1 \to 0$ | 6.403 | 6.4971 | | | | |
| $0 \to 1$ | 7.044 | 7.2396 | | | | |

**Table 5.** Out-of-Phase AC Magnetic Susceptibility Results for **(2)**.

| | [Mn$_4$(hmp)$_6$Br$_2$(H$_2$O)$_2$]Br$_2$·4H$_2$O **(2)** | | |
|---|---|---|---|
| AC Frequency (Hz) | Peak Temperature [K][a] | ln (1/$\tau$)[b] | 1/$T$ [K] |
| 10.5 | 1.044 | -2.3514 | 0.9579 |
| 41 | 1.170 | -3.7136 | 0.8547 |
| 171 | 1.298 | -5.1417 | 0.7704 |
| 241 | 1.344 | -5.4848 | 0.7441 |
| 491 | 1.399 | -6.1964 | 0.7143 |
| 991 | 1.501 | -6.8987 | 0.6662 |

[a]This is the temperature where there is a maximum in the out-of-phase AC magnetic susceptibility $\chi_M''$. [b]This is the natural logarithm of the inverse of the magnetization relaxation time calculated from the frequency of the AC magnetic field.

39**Figure Captions**.

Figure 1. Plot of the potential energy *vs*. magnetization direction for a single-molecule magnet showing the energy barrier separating the "spin-up" and "spin-down" orientations of the magnetic vector along the anisotropy axis.

Figure 2. Diagram of the ligands used in this work.

Figure 3. ORTEP representation of the cation of complex **2**, [Mn$_4$(hmp)$_6$Br$_2$(H$_2$O)$_2$]Br$_2 \cdot$4H$_2$O.

Figure 4. ORTEP representation of complex **3**, [Mn$_4$(6me-hmp)$_6$Cl$_4$]$\cdot$4H$_2$O.

Figure 5. View of the crystal packing of the Mn$_4$ molecules of complex **2**. The dark red spheres are the bromide counterions and the light red spheres are the solvate water molecules.

Figure 6. View of the two Mn$_4$ molecules of complex **2** in the unit cell related by a glide plane. The easy molecular axes of these molecules are related by a canting angle of 9.0 $\pm$ 0.5°.

Figure 7. View of the crystal packing of complex **3**. The red spheres are the solvate water molecules within the unit cell.

Figure 8. Plot of $\chi_M T$ *vs*. temperature for a microcrystalline sample of complex **2** restrained in eicosane. The susceptibility, $\chi_M$, was measured under a 10 kG magnetic field with the sample restrained in eicosane. The solid line corresponds to the fit of the data as described in the text.

Figure 9. Diagram showing the two dominant magnetic exchange pathways between metal centers in complex **2**.

Figure 10. Plot of $\chi_M T$ *vs*. temperature for a microcrystalline sample of complex **3** restrained in eicosane. The susceptibility, $\chi_M$, was measured under a 10 kG magnetic field with the sample restrained in eicosane. The solid line corresponds to the fit of the data as described in the text.

Figure 11. Diagram showing the dominant magnetic exchange pathways between metal centers in complex **3**.

Figure 12. Plot of the reduced magnetizaton *M/N$\mu_B$* *vs*. the ratio of external field and the absolute temperature [kG/K]. Data for complex **2** (top) and complex **3** (bottom) were measured at 20 (■), 30 (●), 40 (▲), and 50 (▼) kG in the 2-4 K range. The solid lines correspond to the fit (see text).

Figure 13. Plot of the energy *vs*. the external field (*H*) for the 19 zero-field split components of the *S* = 9 ground state assuming $D/k_B$ = -0.5 K and *g* = 1.99. The energy level splitting is for a molecule with its principle magnetic axis aligned along the direction of applied magnetic field.

Figure 14. High-frequency EPR spectra for a oriented microcrystalline sample of complex **2** collected at 291.26 GHz and temperatures of 30 and 60 K. The astericks denote the fine structure resonances due to the transitions between Ms levels of the ground state.

Figure 15. High-frequency EPR spectra for a microcrystalline sample of complex **3** collected at 190.01 and 285.02 GHz and temperature of 60 K.

Figure 16. High-frequency EPR spectra for a microcrystalline sample of complex **2** collected at 191.30, 291.26, and 388.35 GHz and temperature of 60 K.

Figure 17. Plots of resonance field *vs*. $M_S$ number for the HFEPR transitions between the $M_S$ and $M_S$+1 zero-field components of the *S* = 9 ground state for complex **2**. HFEPR data were measured at 191.30 (lowest line), 291.26 (middle line), and 388.35 (highest line) GHz at 60 K. The solid lines represent a fit of the data as described in the text.

Figure 18. HFEPR spectrum at 191.3 GHz at 60 K (top) and simulation of the spectrum (bottom). Simulation results are described fully in the text.

Figure 19. Simulation of the HFEPR spectra at 291.26, and 388.35 GHz frequencies at 60 K. See text for details.

Figure 20. Plot of $\chi'$ and $\chi''$ vs. temperature of the AC magnetic susceptibility at the indicated frequencies in the 1.8-6.4 K range.

Figure 21. Plots of $\chi_M'T$ (top) and $\chi_M''$ (bottom) vs. temperature for a polycrystalline sample of complex **2** in a 1.0 G AC field oscillating at the indicated frequencies, where $\chi_M'$ and $\chi_M''$ are the in-phase and out-of-phase components, respectively, of the AC magnetic susceptibility.

Figure 22. Plot of the in-phase (top) and out-of-phase (bottom) components of the AC magnetic susceptibility for complex **3** in the 1.8 - 6.0 K range in a 1 G AC field oscillating at 1000 Hz.

Figure 23. Plot of natural logarithm of the magnetization relaxation rate [ln(1/$\tau$)] *vs*. the inverse of the absolute temperature [K$^{-1}$]. The solid line represents a least-squares fit of the data to the Arrhenius equation (see text).

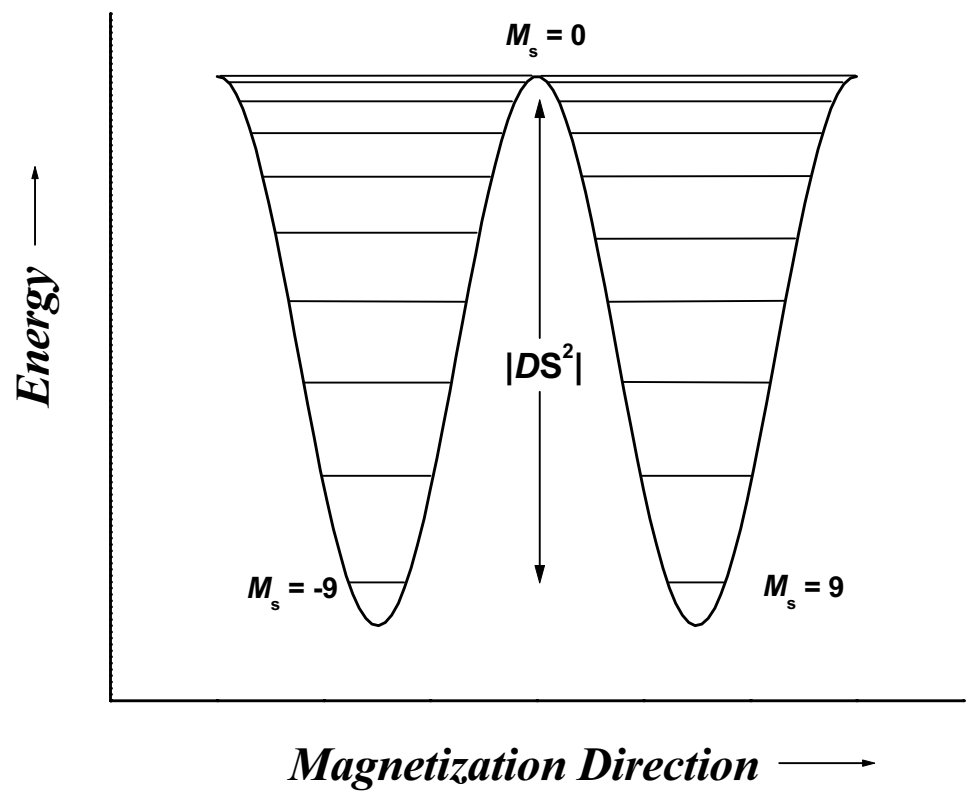

**Figure 1**.

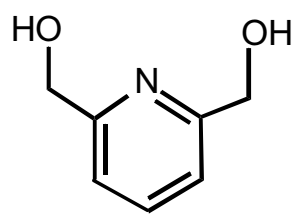

pdmH$_2$

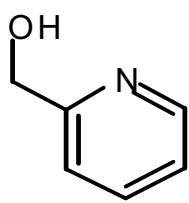

hmpH

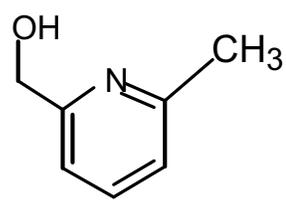

6Me-hmpH

**Figure 2**

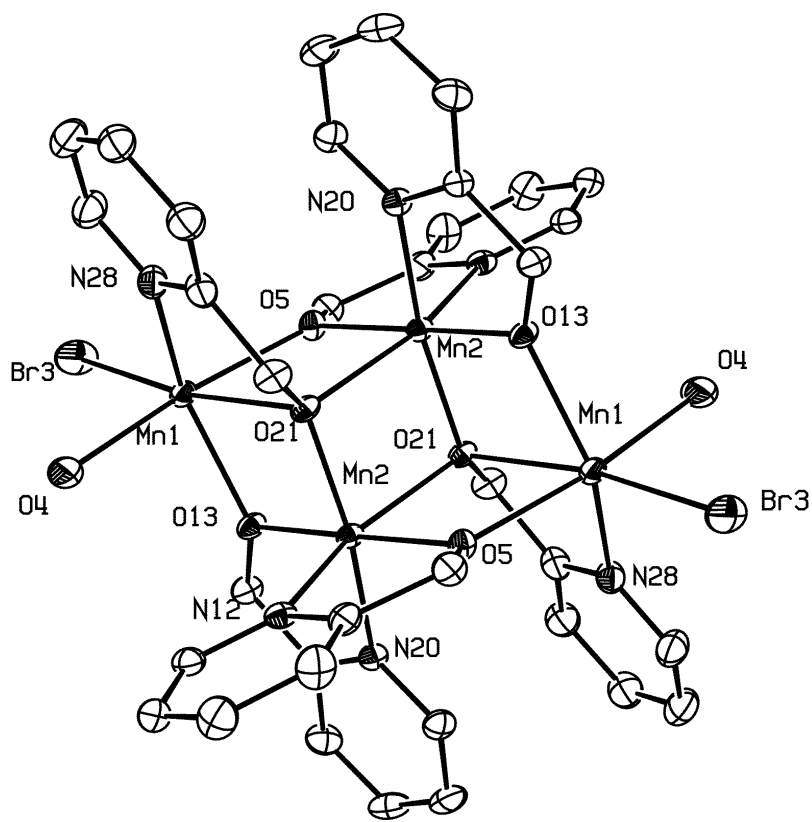

**Figure 3**

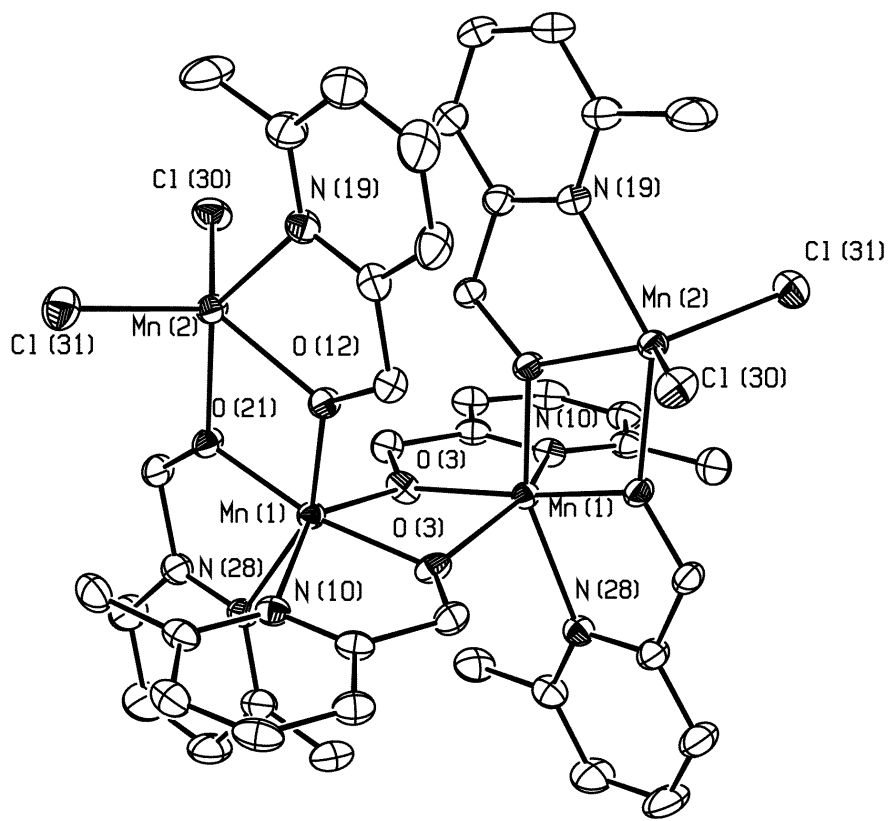

**Figure 4**

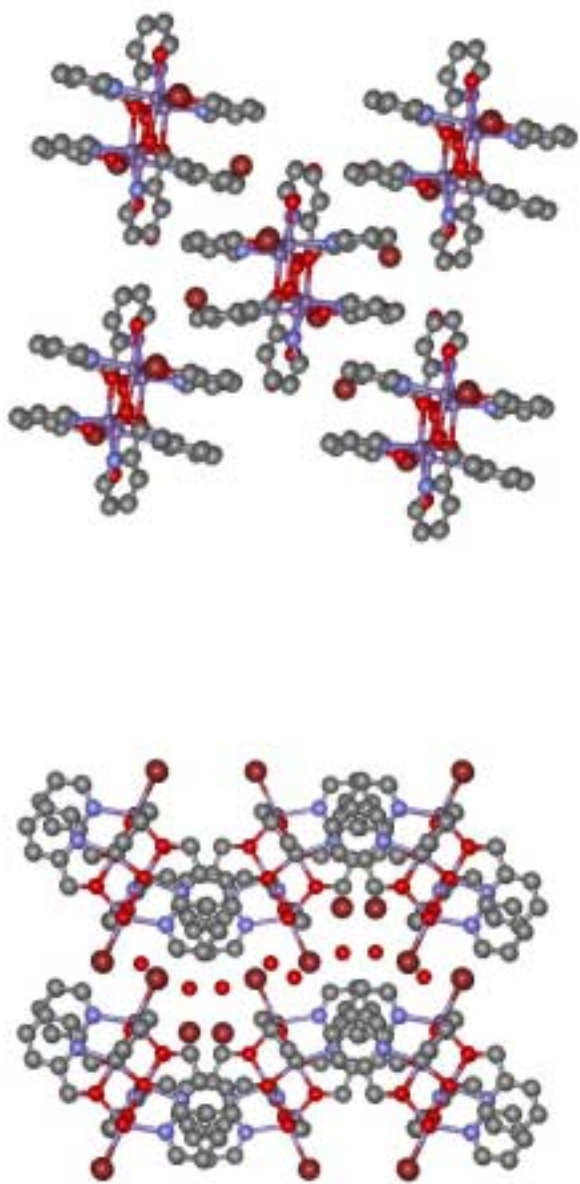

**Figure 5.** View of the crystal packing of the Mn$_4$ molecules of complex **2**. The dark red spheres are the bromide counterions and the light red spheres are the solvate water molecules.

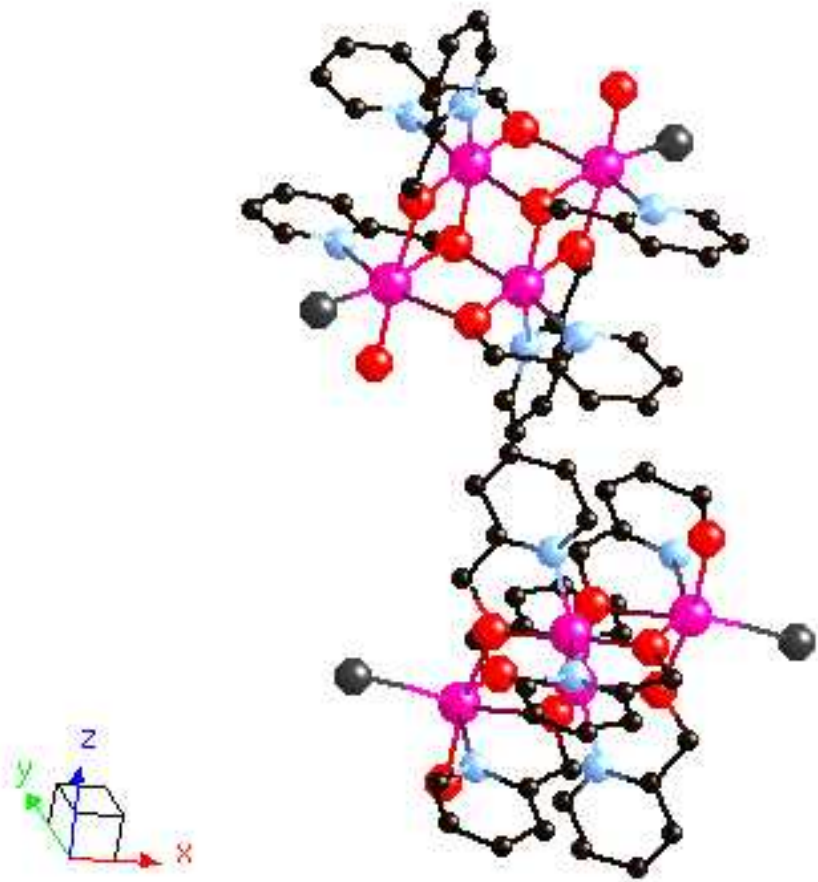

**Figure 6.** View of the two Mn$_4$ molecules of complex **2** in the unit cell related by a glide plane. The easy molecular axes of these molecules are related by a canting angle of 9.0 ± 0.5°.

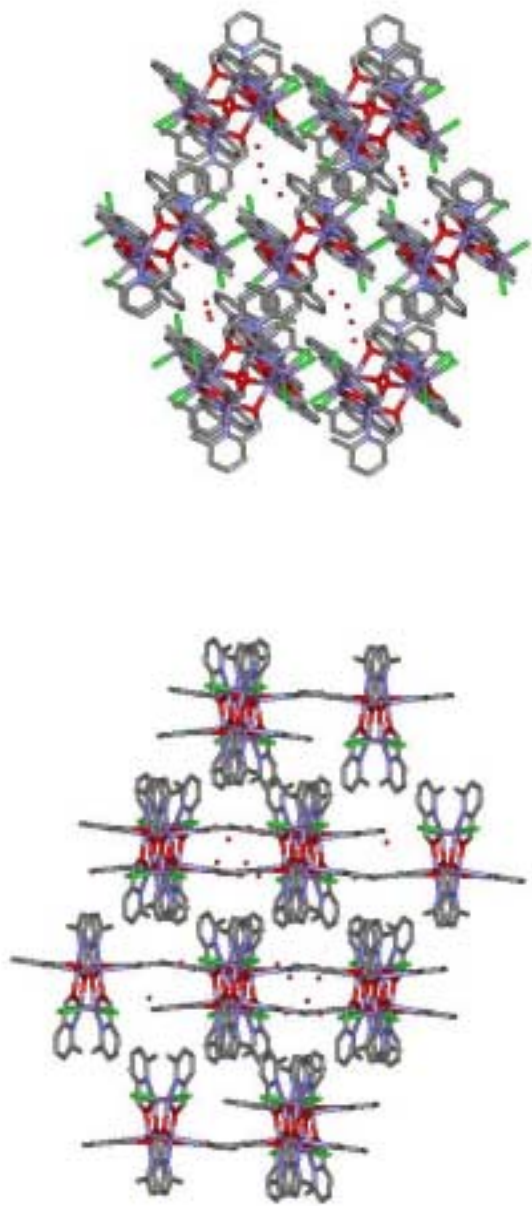

**Figure 7.** View of the crystal packing of complex **3**. The red spheres are the solvate water molecules within the unit cell.

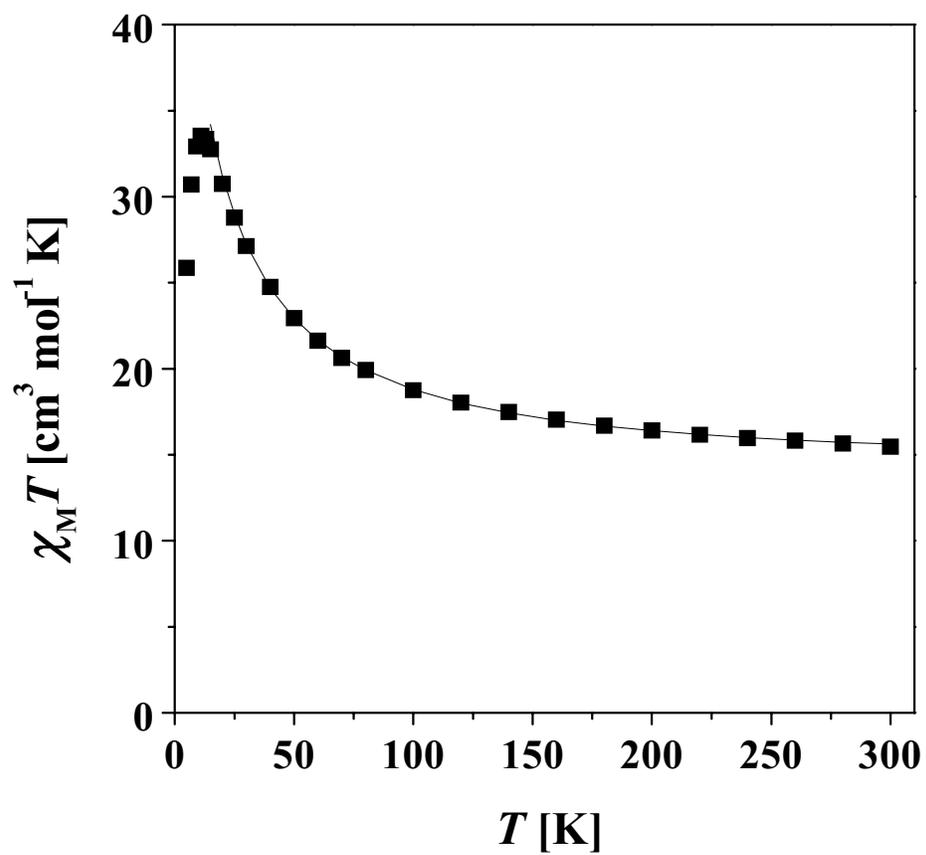

**Figure 8.**

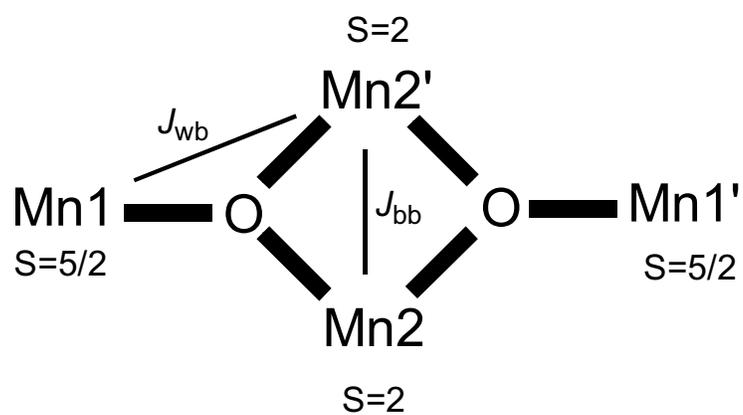

**Figure 9**.

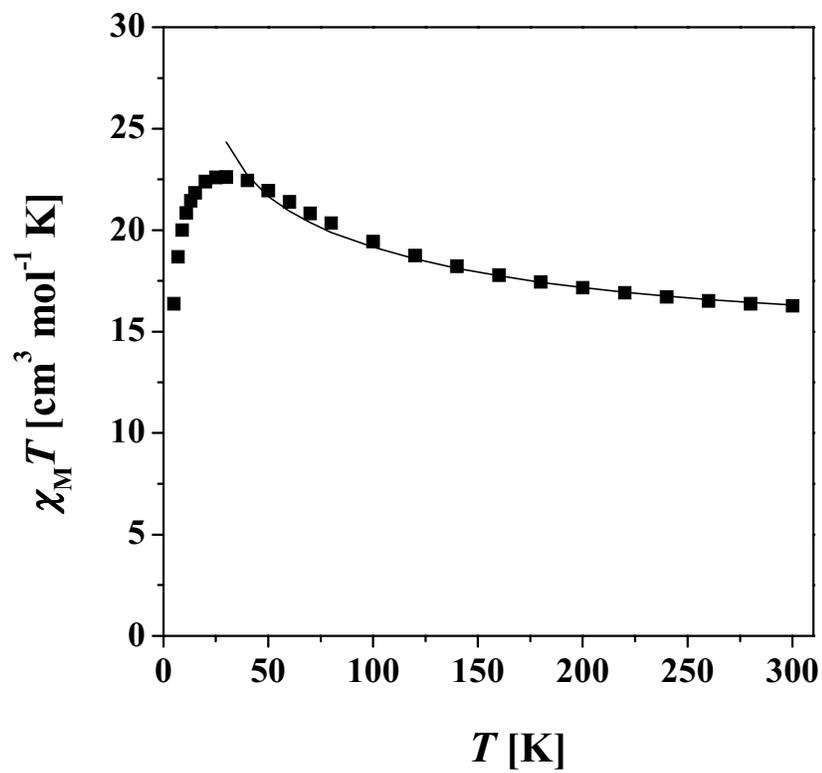

**Figure 10.**

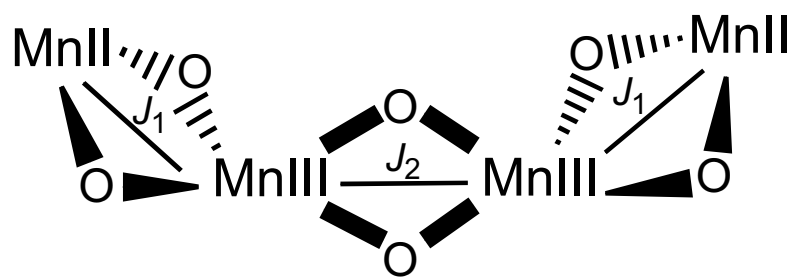

**Figure 11**.

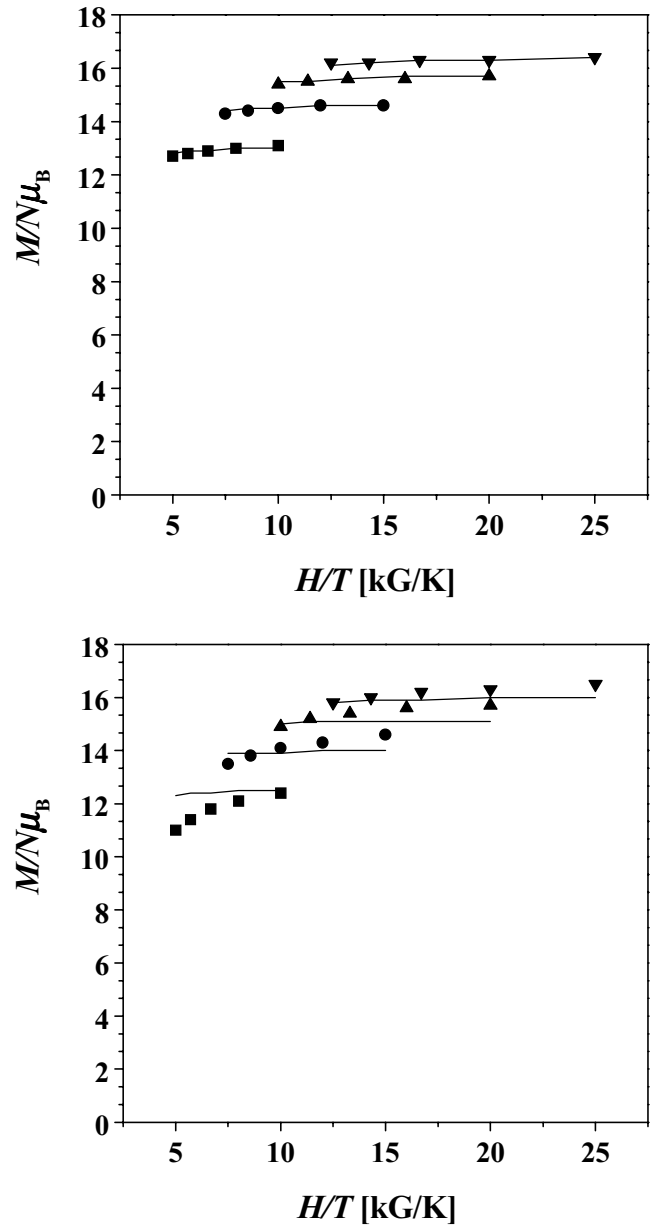

**Figure 12**.

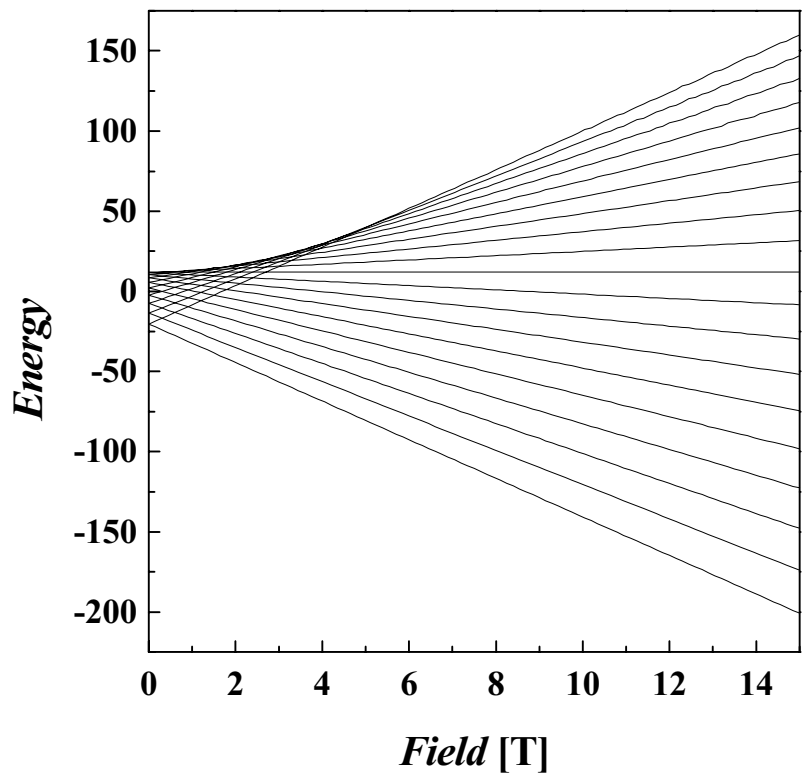

**Figure 13**.

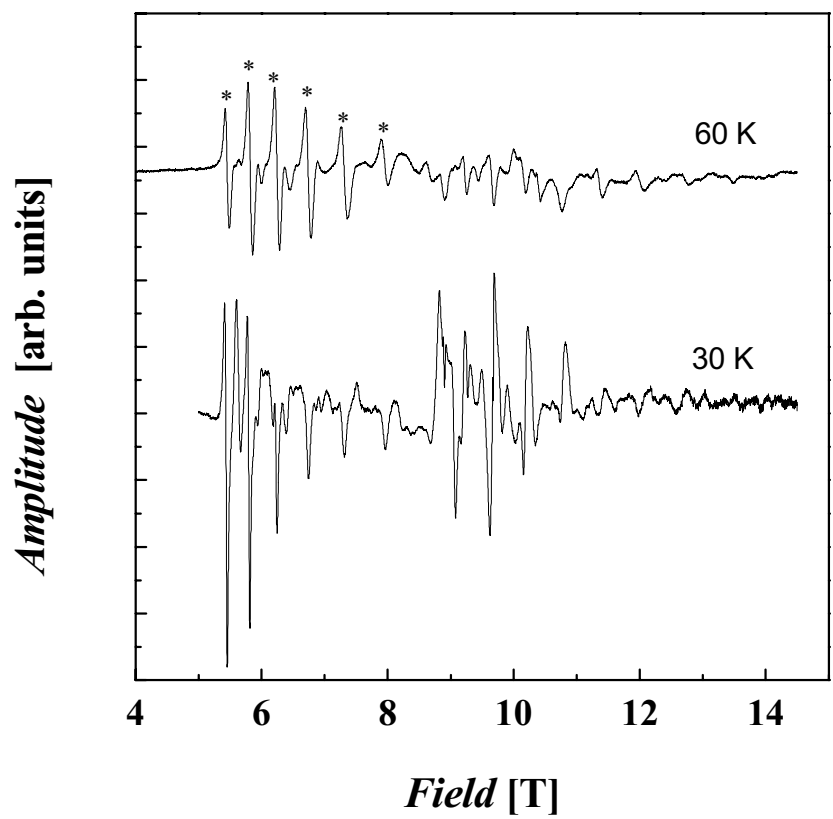

**Figure 14**.

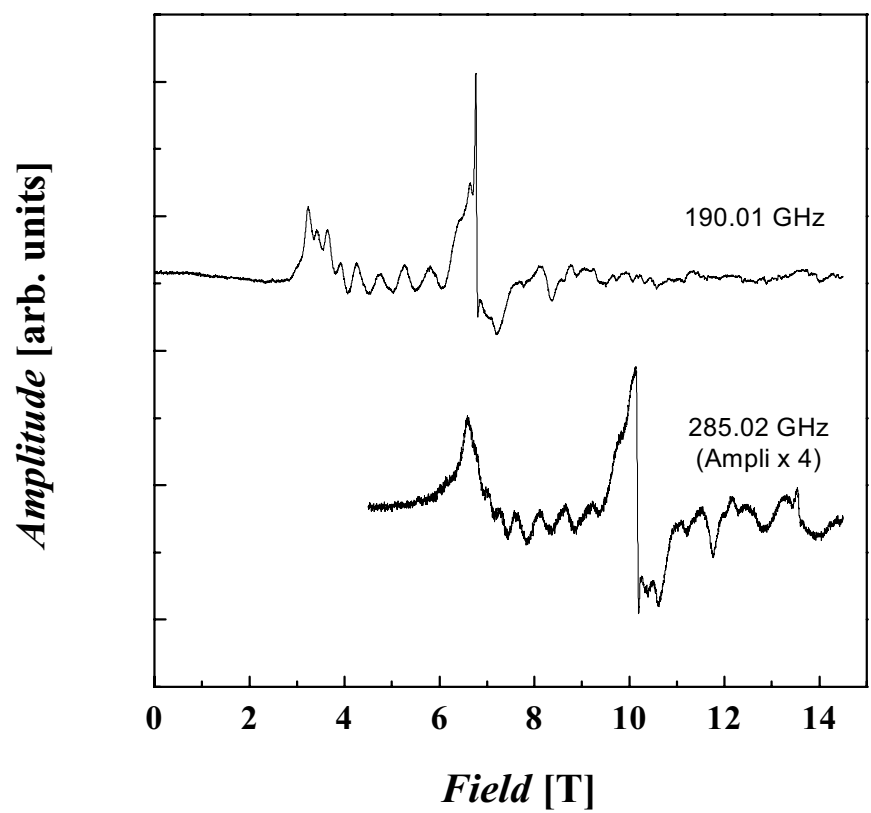

**Figure 15**.

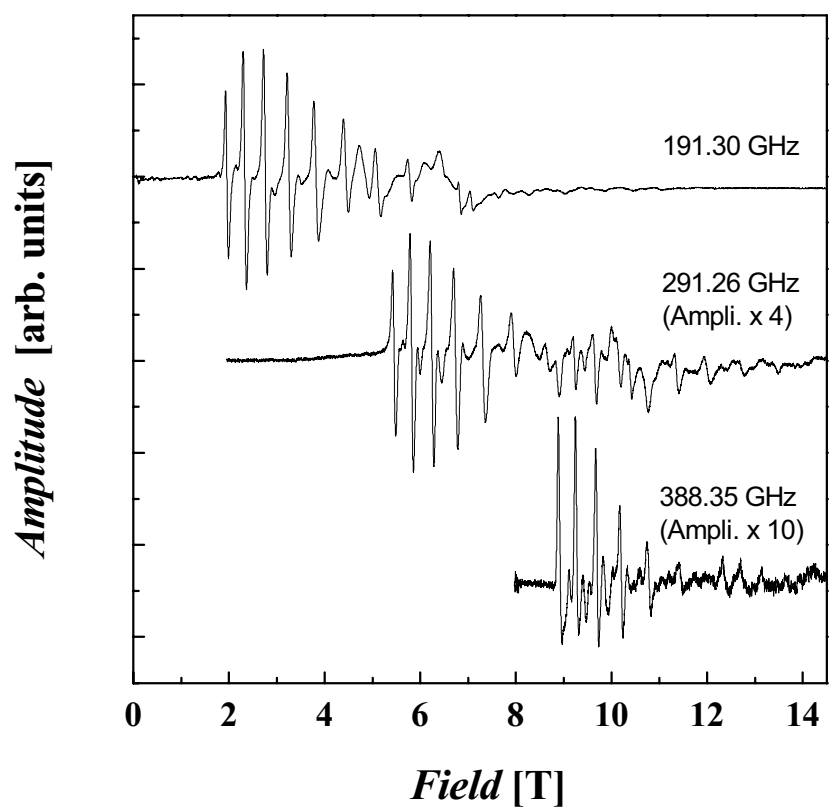

**Figure 16**.

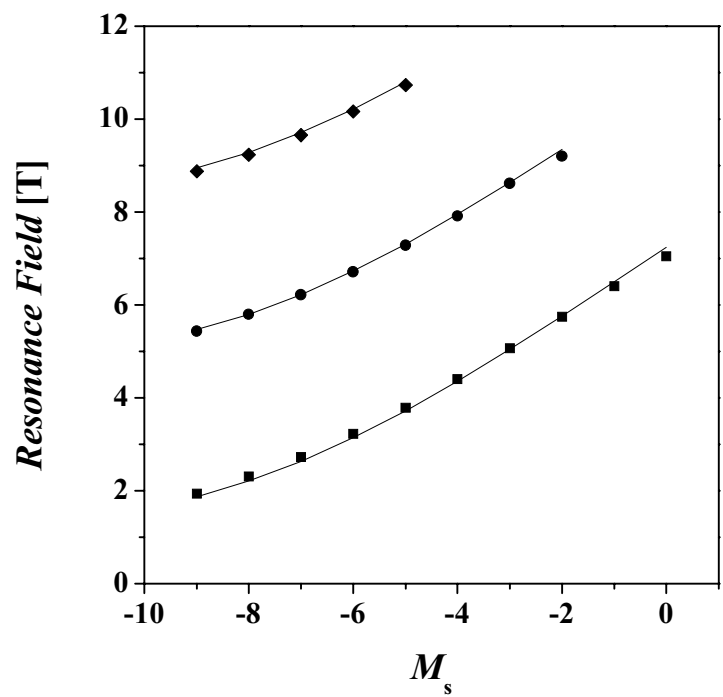

**Figure 17**.

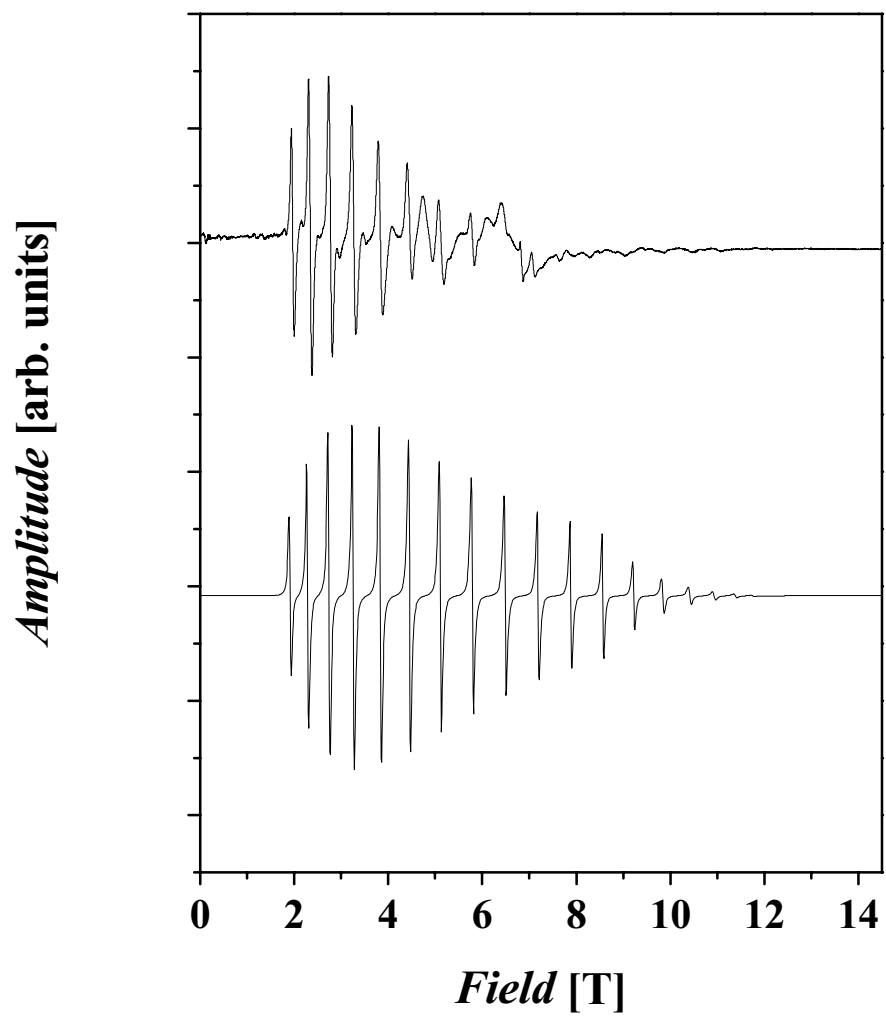

**Figure 18**.

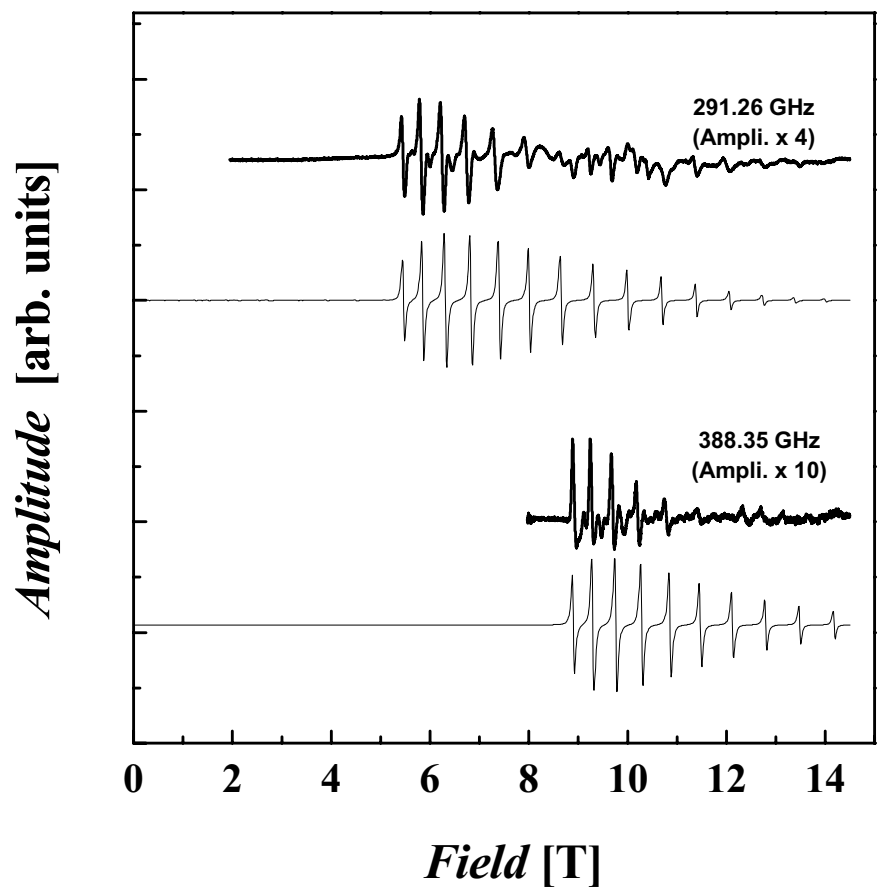

**Figure 19**.

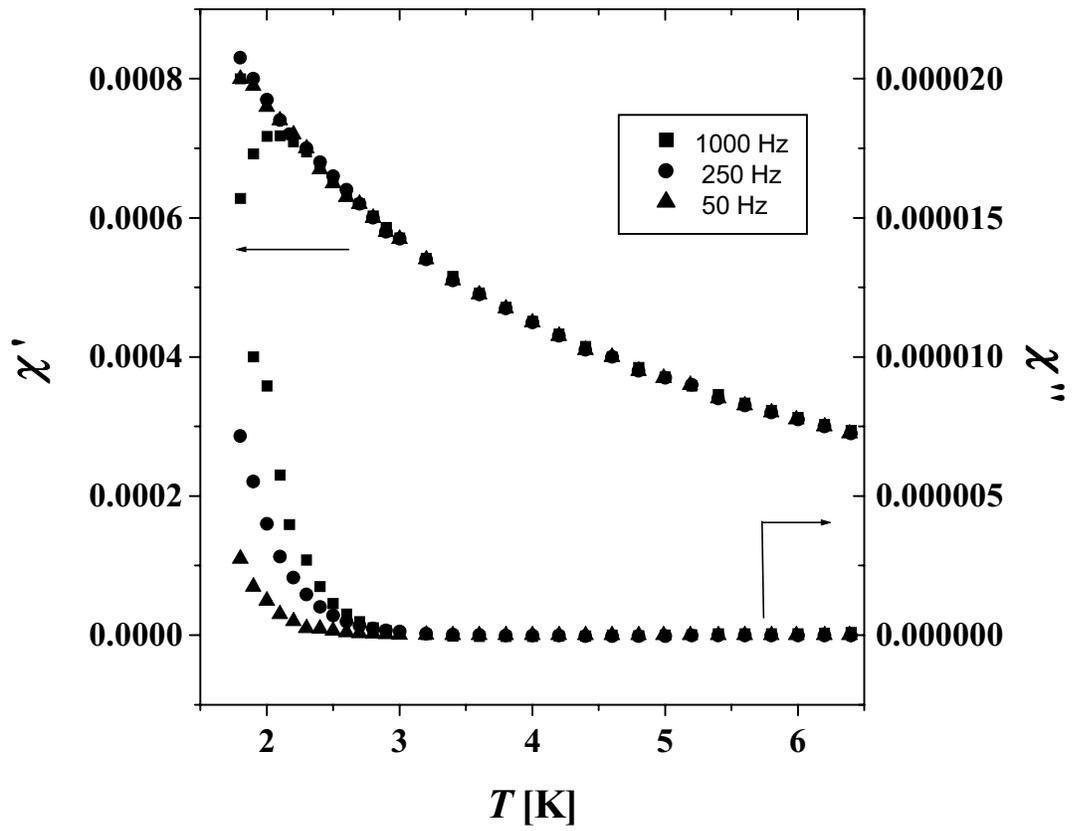

**Figure 20**.

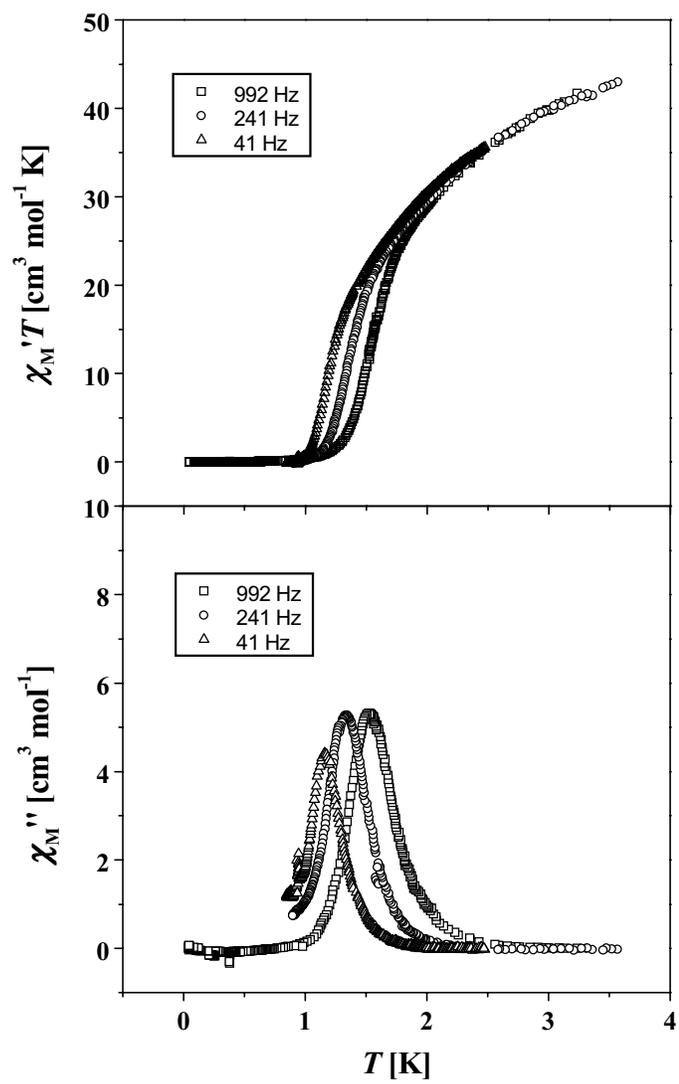

**Figure 21**.

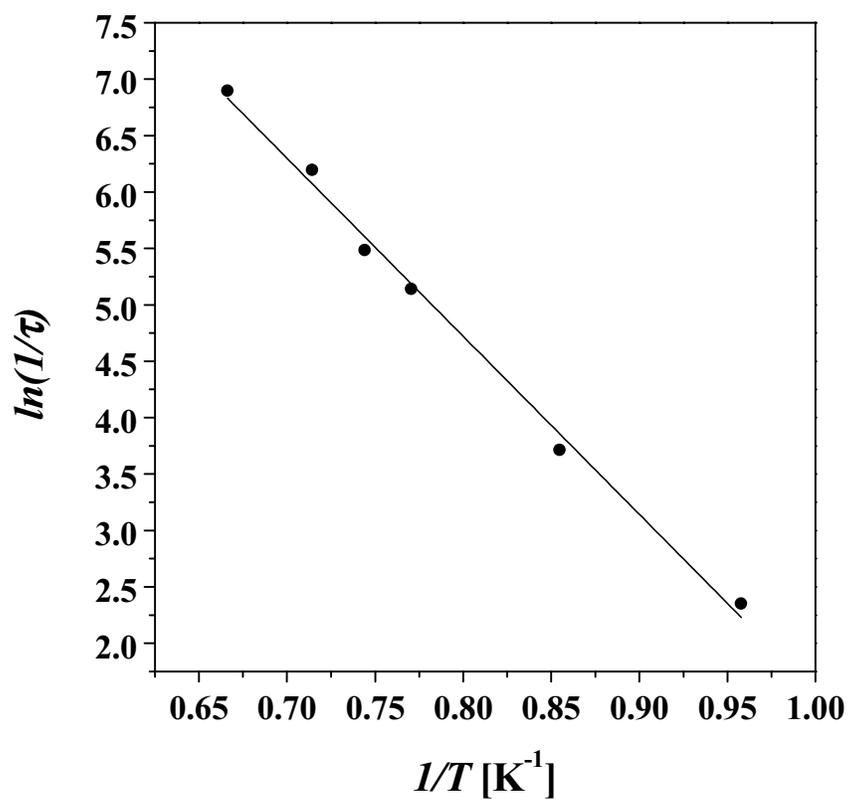

**Figure 22**.

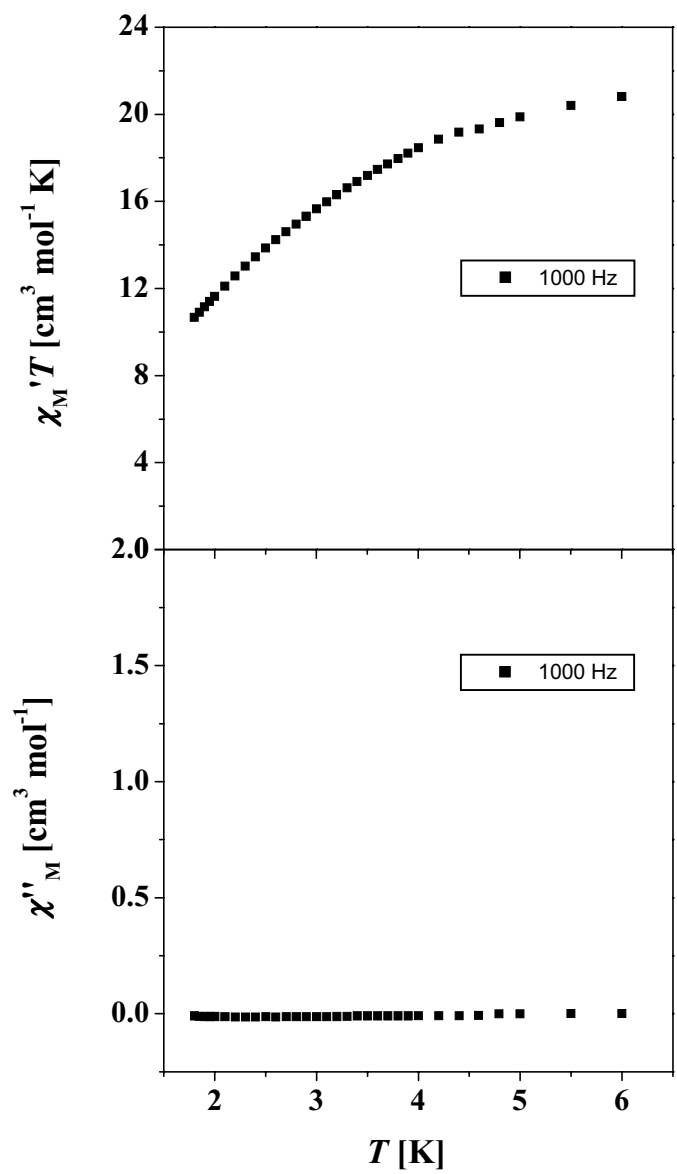

**Figure 23.**

**Appendix**

Appendix 3.1. Atomic coordinates (x,y,z) for complex **2** at ? K.

| | | | | | | | |
|---|---|---|---|---|---|---|---|
| Mn1 | 0.75364(4) | 0.01918(3) | 0.87874(3) | O30 | 0.4124(2) | -0.19528(18) | 0.5911(2) |
| Mn2 | 1.05353(4) | 0.04877(3) | 0.91664(3) | O31 | 0.5314(3) | -0.0373(2) | 0.6025(3) |
| Br3 | 0.53930(3) | 0.08853(2) | 0.86418(3) | H1 | 0.761(3) | 0.062(3) | 0.693(3) |
| O4 | 0.7497(2) | 0.01951(17) | 0.71848(17) | H2 | 0.689(3) | 0.005(2) | 0.691(2) |
| O5 | 0.79294(15) | 0.00710(11) | 1.03865(13) | H3 | 0.751(2) | 0.1164(19) | 1.090(2) |
| C6 | 0.7381(3) | 0.0575(2) | 1.1030(2) | H4 | 0.657(3) | 0.0449(18) | 1.090(2) |
| C7 | 0.7887(2) | 0.03452(17) | 1.2081(2) | H5 | 0.671(3) | 0.1019(19) | 1.270(2) |
| C8 | 0.7375(3) | 0.0643(2) | 1.2850(2) | H6 | 0.759(3) | 0.053(2) | 1.435(2) |
| C9 | 0.7855(3) | 0.0363(2) | 1.3787(2) | H7 | 0.919(3) | -0.0417(19) | 1.456(2) |
| C10 | 0.8852(3) | -0.0202(2) | 1.3933(2) | H8 | 0.999(2) | -0.0762(17) | 1.3211(18) |
| C11 | 0.9356(3) | -0.04317(18) | 1.3146(2) | H9 | 0.851(2) | 0.2206(16) | 0.8127(19) |
| N12 | 0.88845(19) | -0.01701(14) | 1.22291(16) | H10 | 0.865(2) | 0.2183(17) | 0.9347(19) |
| O13 | 0.90273(16) | 0.10871(11) | 0.88246(13) | H11 | 1.007(3) | 0.355(2) | 0.879(2) |
| C14 | 0.9026(3) | 0.19846(18) | 0.8816(2) | H12 | 1.212(3) | 0.393(2) | 0.901(2) |
| C15 | 1.0353(2) | 0.22961(17) | 0.89587(19) | H13 | 1.367(3) | 0.2869(18) | 0.932(2) |
| C16 | 1.0670(3) | 0.31461(19) | 0.8925(2) | H14 | 1.301(2) | 0.1455(17) | 0.9350(19) |
| C17 | 1.1919(3) | 0.3362(2) | 0.9061(2) | H15 | 0.963(2) | -0.1150(18) | 0.800(2) |
| C18 | 1.2814(3) | 0.2736(2) | 0.9227(2) | H16 | 0.996(3) | -0.174(2) | 0.897(2) |
| C19 | 1.2453(3) | 0.1908(2) | 0.9255(2) | H17 | 0.891(3) | -0.2885(19) | 0.836(2) |
| N20 | 1.12318(19) | 0.16894(14) | 0.91326(16) | H18 | 0.679(3) | -0.350(2) | 0.803(2) |
| O21 | 0.94304(16) | -0.04803(11) | 0.92101(13) | H19 | 0.503(3) | -0.2689(18) | 0.823(2) |
| C22 | 0.9451(3) | -0.12482(19) | 0.8689(2) | H20 | 0.539(3) | -0.122(2) | 0.854(2) |
| C23 | 0.8190(3) | -0.16840(19) | 0.8532(2) | H21 | 0.416(3) | -0.235(2) | 0.612(3) |
| C24 | 0.8080(3) | -0.2549(2) | 0.8362(2) | H22 | 0.362(4) | -0.198(3) | 0.520(3) |
| C25 | 0.6895(3) | -0.2913(2) | 0.8205(2) | H23 | 0.522(3) | -0.081(2) | 0.614(3) |
| C26 | 0.5857(3) | -0.2413(2) | 0.8230(2) | H24 | 0.508(5) | -0.035(3) | 0.549(3) |
| C27 | 0.6044(3) | -0.1555(2) | 0.8401(2) | | | | |
| N28 | 0.7173(2) | -0.11929(15) | 0.85477(17) | | | | |
| Br29 | 0.80097(3) | 0.19619(2) | 0.60897(2) | | | | |

Appendix 3.2. Bond distances for complex **2**.

| | | | | | | |
|---|---|---|---|---|---|---|
| Mn1 | O4 | 2.226(2) | | Mn2 | N12 | 2.217(2) |
| Mn1 | O5 | 2.1950(18) | | Mn2 | N20 | 2.048(2) |
| Mn1 | O13 | 2.1477(18) | | O4 | H1 | 0.78(4) |
| Mn1 | O21 | 2.2952(17) | | O4 | H2 | 0.73(3) |
| Mn1 | N28 | 2.235(2) | | O5 | C6 | 1.416(3) |
| Mn2 | O5 | 1.8882(17) | | O13 | C14 | 1.417(3) |
| Mn2 | O13 | 1.8770(17) | | O21 | C22 | 1.416(3) |
| Mn2 | O21 | 1.9545(18) | | O30 | H21 | 0.69(3) |
| Mn2 | O21 | 2.2563(18) | | O30 | H22 | 1.04(4) |
| Mn2 | O21 | 2.2563(18) | | O31 | H23 | 0.72(3) |

| | | | | | | | |
|---|---|---|---|---|---|---|---|
| O31 | H24 | 0.73(4) | | C14 | H10 | 0.96(3) | |
| N12 | C7 | 1.342(3) | | C15 | C16 | 1.389(4) | |
| N12 | C11 | 1.346(3) | | C16 | C17 | 1.381(4) | |
| N20 | C15 | 1.344(3) | | C16 | H11 | 0.91(3) | |
| N20 | C19 | 1.355(3) | | C17 | C18 | 1.377(4) | |
| N28 | C23 | 1.357(3) | | C17 | H12 | 0.94(3) | |
| N28 | C27 | 1.337(4) | | C18 | C19 | 1.368(4) | |
| C6 | C7 | 1.507(4) | | C18 | H13 | 0.94(3) | |
| C6 | H3 | 0.96(3) | | C19 | H14 | 0.93(3) | |
| C6 | H4 | 0.89(3) | | C22 | C23 | 1.515(4) | |
| C7 | C8 | 1.384(4) | | C22 | H15 | 1.02(3) | |
| C8 | C9 | 1.382(4) | | C22 | H16 | 0.99(3) | |
| C8 | H5 | 0.93(3) | | C23 | C24 | 1.387(4) | |
| C9 | C10 | 1.391(4) | | C24 | C25 | 1.393(4) | |
| C9 | H6 | 0.92(3) | | C24 | H17 | 1.05(3) | |
| C10 | C11 | 1.368(4) | | C25 | C26 | 1.386(5) | |
| C10 | H7 | 0.94(3) | | C25 | H18 | 0.96(3) | |
| C11 | H8 | 0.86(3) | | C26 | C27 | 1.385(4) | |
| C14 | C15 | 1.504(4) | | C26 | H19 | 1.01(3) | |
| C14 | H9 | 1.07(3) | | C27 | H20 | 0.94(3) | |

Appendix 3.3.  Bond Angles for complex **2**.

| | | | | | | | |
|---|---|---|---|---|---|---|---|
| O4 | Mn1 | O5 | 168.89(8) | O21 | Mn2 | N20 | 95.39(8) |
| O4 | Mn1 | O13 | 83.66(9) | N12 | Mn2 | N20 | 91.08(8) |
| O4 | Mn1 | O21 | 95.71(8) | Mn1 | O4 | H1 | 120(3) |
| O4 | Mn1 | N28 | 83.33(9) | Mn1 | O4 | H2 | 111(3) |
| O5 | Mn1 | O13 | 92.02(7) | H1 | O4 | H2 | 103(4) |
| O5 | Mn1 | O21 | 73.18(6) | Mn1 | O5 | Mn2 | 111.19(8) |
| O5 | Mn1 | N28 | 93.48(8) | Mn1 | O5 | C6 | 124.86(16) |
| O13 | Mn1 | O21 | 70.03(6) | Mn2 | O5 | C6 | 120.68(16) |
| O13 | Mn1 | N28 | 139.34(8) | Mn1 | O13 | Mn2 | 107.24(8) |
| O21 | Mn1 | N28 | 73.18(7) | Mn1 | O13 | C14 | 131.19(15) |
| O5 | Mn2 | O13 | 175.21(8) | Mn2 | O13 | C14 | 120.36(15) |
| O5 | Mn2 | O21 | 97.88(8) | Mn1 | O21 | Mn2 | 95.66(6) |
| O5 | Mn2 | O21 | 79.96(7) | Mn1 | O21 | Mn2 | 99.22(8) |
| O5 | Mn2 | N12 | 78.34(8) | Mn1 | O21 | C22 | 111.55(15) |
| O5 | Mn2 | N20 | 97.54(8) | Mn2 | O21 | Mn2 | 98.35(7) |
| O13 | Mn2 | O21 | 83.51(8) | Mn2 | O21 | C22 | 120.77(17) |
| O13 | Mn2 | O21 | 95.75(7) | Mn2 | O21 | C22 | 125.60(16) |
| O13 | Mn2 | N12 | 106.07(8) | H21 | O30 | H22 | 110(4) |
| O13 | Mn2 | N20 | 80.66(8) | H23 | O31 | H24 | 104(5) |
| O21 | Mn2 | O21 | 81.65(7) | Mn2 | N12 | C7 | 111.32(17) |
| O21 | Mn2 | N12 | 97.67(8) | Mn2 | N12 | C11 | 129.94(19) |
| O21 | Mn2 | N12 | 157.99(7) | C7 | N12 | C11 | 118.7(2) |
| O21 | Mn2 | N20 | 163.53(8) | Mn2 | N20 | C15 | 114.19(17) |
| Mn2 | N20 | C19 | 126.42(19) | C15 | C16 | H11 | 121(2) |
| C15 | N20 | C19 | 119.4(2) | C17 | C16 | H11 | 120(2) |
| Mn1 | N28 | C23 | 116.08(18) | C16 | C17 | C18 | 119.7(3) |
| Mn1 | N28 | C27 | 124.9(2) | C16 | C17 | H12 | 118.0(19) |
| C23 | N28 | C27 | 119.0(3) | C18 | C17 | H12 | 122.2(19) |
| O5 | C6 | C7 | 111.2(2) | C17 | C18 | C19 | 119.4(3) |
| O5 | C6 | H3 | 109.3(17) | C17 | C18 | H13 | 121.0(18) |
| O5 | C6 | H4 | 105.4(19) | C19 | C18 | H13 | 119.6(18) |
| C7 | C6 | H3 | 112.2(17) | N20 | C19 | C18 | 121.4(3) |
| C7 | C6 | H4 | 108.3(18) | N20 | C19 | H14 | 114.7(16) |
| H3 | C6 | H4 | 110(2) | C18 | C19 | H14 | 123.9(16) |
| N12 | C7 | C6 | 115.4(2) | O21 | C22 | C23 | 111.0(2) |
| N12 | C7 | C8 | 121.6(3) | O21 | C22 | H15 | 112.0(17) |
| C6 | C7 | C8 | 123.0(3) | O21 | C22 | H16 | 121.7(18) |
| C7 | C8 | C9 | 119.2(3) | C23 | C22 | H15 | 105.5(15) |
| C7 | C8 | H5 | 116.9(18) | C23 | C22 | H16 | 97.2(18) |
| C9 | C8 | H5 | 123.9(18) | H15 | C22 | H16 | 107(2) |
| C8 | C9 | C10 | 119.0(3) | N28 | C23 | C22 | 117.4(3) |
| C8 | C9 | H6 | 126(2) | N28 | C23 | C24 | 121.4(3) |
| C10 | C9 | H6 | 115(2) | C22 | C23 | C24 | 121.2(3) |
| C9 | C10 | C11 | 118.5(3) | C23 | C24 | C25 | 118.7(3) |
| C9 | C10 | H7 | 121.2(18) | C23 | C24 | H17 | 117.0(16) |
| C11 | C10 | H7 | 120.3(18) | C25 | C24 | H17 | 124.3(16) |
| N12 | C11 | C10 | 122.8(3) | C24 | C25 | C26 | 119.9(3) |
| N12 | C11 | H8 | 116.1(18) | C24 | C25 | H18 | 119.7(19) |
| C10 | C11 | H8 | 121.1(18) | C26 | C25 | H18 | 120.2(18) |
| O13 | C14 | C15 | 109.1(2) | C25 | C26 | C27 | 117.9(3) |
| O13 | C14 | H9 | 109.4(14) | C25 | C26 | H19 | 119.7(16) |
| O13 | C14 | H10 | 108.5(16) | C27 | C26 | H19 | 121.7(16) |
| C15 | C14 | H9 | 109.7(14) | N28 | C27 | C26 | 123.1(3) |
| C15 | C14 | H10 | 109.5(15) | N28 | C27 | H20 | 116.1(19) |
| H9 | C14 | H10 | 111(2) | C26 | C27 | H20 | 119.7(19) |
| N20 | C15 | C14 | 115.2(2) | | | | |
| N20 | C15 | C16 | 121.4(3) | | | | |
| C14 | C15 | C16 | 123.4(3) | | | | |
| C15 | C16 | C17 | 118.6(3) | | | | |

Appendix 3.4.  Atomic coordinates for complex **3**.

Mn1 0.09347(3)   0.30244(2)    0.74237(3)  C8  0.2799(2)    0.33333(16) 0.9715(3)
Mn2 0.16081(3)   0.16955(2)    0.65721(3)  C9  0.24598(19) 0.32441(15) 0.8830(2)
O3   0.01623(13) 0.31904(11)  0.83557(1)   N10 0.16695(16) 0.31984(12) 0.87353(18)
C4   0.0345(2)    0.32029(17) 0.9325(2)    C11 0.2940(2)    0.3189(2)    0.7963(3)
C5   0.1222(2)    0.32317(15) 0.9498(2)    O12 0.09499(13) 0.21176(10) 0.76423(15)
C6   0.1530(2)    0.33057(16) 1.0414(2)    C13 0.0669(2)    0.17739(16) 0.8416(2)
C7   0.2331(2)    0.33567(16) 1.0506(2)    C14 0.08224(19) 0.10721(16) 0.8300(2)

| | | | | | | |
|---|---|---|---|---|---|---|
| C15 | 0.0597(2) | 0.06445(19) | 0.8992(3) | H4 | 0.253(2) | 0.3447(19) | 1.109(3) |
| C16 | 0.0731(3) | 0.0002(2) | 0.8863(3) | H5 | 0.338(2) | 0.3383(17) | 0.968(3) |
| C17 | 0.1082(2) | -0.01999(18) | 0.8052(3) | H6 | 0.288(2) | 0.283(2) | 0.773(3) |
| C18 | 0.1304(2) | 0.02433(17) | 0.7374(3) | H7 | 0.279(3) | 0.351(2) | 0.753(3) |
| N19 | 0.11735(16) | 0.08783(13) | 0.75057(1) | H8 | 0.348(2) | 0.3203(18) | 0.811(3) |
| C20 | 0.1704(3) | 0.0050(2) | 0.6481(3) | H9 | 0.091(2) | 0.1887(16) | 0.887(2) |
| O21 | 0.15508(13) | 0.27360(10) | 0.64301(1) | H10 | 0.011(2) | 0.1875(18) | 0.851(3) |
| C22 | 0.1761(2) | 0.31624(16) | 0.5707(3) | H11 | 0.033(2) | 0.081(2) | 0.943(3) |
| C23 | 0.16491(19) | 0.38416(16) | 0.6013(2) | H12 | 0.054(3) | -0.028(2) | 0.926(3) |
| C24 | 0.1918(2) | 0.43472(18) | 0.5481(3) | H13 | 0.121(2) | -0.064(2) | 0.802(3) |
| C25 | 0.1832(3) | 0.49579(19) | 0.5820(3) | H14 | 0.163(3) | -0.029(3) | 0.631(4) |
| C26 | 0.1498(2) | 0.50528(18) | 0.6693(3) | H15 | 0.142(2) | 0.0229(18) | 0.594(3) |
| C27 | 0.1236(2) | 0.45328(16) | 0.7209(2) | H16 | 0.227(3) | 0.015(2) | 0.649(3) |
| N28 | 0.13091(16) | 0.39274(13) | 0.68625(1) | H17 | 0.223(2) | 0.3106(16) | 0.559(2) |
| C29 | 0.0884(3) | 0.46246(19) | 0.8161(3) | H18 | 0.150(2) | 0.310(2) | 0.526(3) |
| CL30 | 0.11535(5) | 0.14820(4) | 0.50135(6) | H19 | 0.220(2) | 0.426(2) | 0.495(3) |
| CL31 | 0.29598(5) | 0.14928(4) | 0.68486(6) | H20 | 0.202(2) | 0.528(2) | 0.553(3) |
| O32 | -0.0069(2) | 0.2063(3) | 0.3504(3) | H21 | 0.148(2) | 0.5480(18) | 0.699(3) |
| O33 | 0.4786(2) | 0.1631(2) | 0.6558(2) | H22 | 0.075(2) | 0.503(2) | 0.825(3) |
| H1 | 0.011(2) | 0.3581(17) | 0.961(2) | H23 | 0.047(2) | 0.4386(19) | 0.834(3) |
| H2 | 0.014(2) | 0.2824(17) | 0.958(2) | H24 | 0.119(2) | 0.452(2) | 0.856(3) |
| H3 | 0.1183(17) | 0.3297(14) | 1.095(2) | | | | |

Appendix 3.5.  Bond distances for complex **3**.

| | | | | | | |
|---|---|---|---|---|---|---|
| Mn1 | O3 | 1.909(2) | | C4 | H1 | 0.98(3) |
| Mn1 | O3 | 2.194(2) | | C4 | H2 | 0.94(4) |
| Mn1 | O12 | 1.919(2) | | C5 | C6 | 1.405(4) |
| Mn1 | O21 | 1.867(2) | | C6 | C7 | 1.379(5) |
| Mn1 | N10 | 2.264(3) | | C6 | H3 | 0.97(3) |
| Mn1 | N28 | 2.146(3) | | C7 | C8 | 1.385(5) |
| Mn2 | CL30 | 2.3782(9) | | C7 | H4 | 0.92(4) |
| Mn2 | CL31 | 2.3768(10) | | C8 | C9 | 1.390(5) |
| Mn2 | O12 | 2.092(2) | | C8 | H5 | 1.01(4) |
| Mn2 | O21 | 2.184(2) | | C9 | C11 | 1.490(5) |
| Mn2 | N19 | 2.288(3) | | C11 | H6 | 0.82(4) |
| O3 | C4 | 1.407(4) | | C11 | H7 | 0.95(4) |
| O12 | C13 | 1.400(4) | | C11 | H8 | 0.94(4) |
| O21 | C22 | 1.407(4) | | C13 | C14 | 1.498(5) |
| N10 | C5 | 1.332(4) | | C13 | H9 | 0.79(3) |
| N10 | C9 | 1.359(4) | | C13 | H10 | 0.99(4) |
| N19 | C14 | 1.344(4) | | C14 | C15 | 1.385(5) |
| N19 | C18 | 1.358(4) | | C15 | C16 | 1.373(6) |
| N28 | C23 | 1.355(4) | | C15 | H11 | 0.84(4) |
| N28 | C27 | 1.362(4) | | C16 | C17 | 1.369(6) |
| C4 | C5 | 1.516(5) | | C16 | H12 | 0.88(4) |

| C17 | C18 | 1.390(5) | C24 | C25 | 1.371(5) |
| C17 | H13 | 0.95(4) | C24 | H19 | 0.91(4) |
| C18 | C20 | 1.500(6) | C25 | C26 | 1.384(6) |
| C20 | H14 | 0.76(5) | C25 | H20 | 0.85(4) |
| C20 | H15 | 0.97(4) | C26 | C27 | 1.385(5) |
| C20 | H16 | 0.99(5) | C26 | H21 | 0.98(4) |
| C22 | C23 | 1.495(5) | C27 | C29 | 1.496(5) |
| C22 | H17 | 0.83(4) | C29 | H22 | 0.88(4) |
| C22 | H18 | 0.79(4) | C29 | H23 | 0.90(4) |
| C23 | C24 | 1.379(5) | C29 | H24 | 0.80(4) |

Appendix 3.6. Bond angles for complex **3**.

| O3 | Mn1 | O3 | 74.18(9) | Mn2 | O21 | C22 | 133.3(2) |
| O3 | Mn1 | O12 | 104.32(9) | Mn1 | N10 | C5 | 110.9(2) |
| O3 | Mn1 | O12 | 94.34(9) | Mn1 | N10 | C9 | 129.4(2) |
| O3 | Mn1 | O21 | 98.95(9) | C5 | N10 | C9 | 119.6(3) |
| O3 | Mn1 | O21 | 168.54(10) | Mn2 | N19 | C14 | 114.2(2) |
| O3 | Mn1 | N10 | 148.54(9) | Mn2 | N19 | C18 | 126.5(2) |
| O3 | Mn1 | N10 | 77.49(9) | C14 | N19 | C18 | 119.0(3) |
| O3 | Mn1 | N28 | 86.02(9) | Mn1 | N28 | C23 | 110.2(2) |
| O3 | Mn1 | N28 | 107.93(10) | Mn1 | N28 | C27 | 130.8(2) |
| O12 | Mn1 | O21 | 78.28(9) | C23 | N28 | C27 | 119.0(3) |
| O12 | Mn1 | N10 | 91.09(9) | O3 | C4 | C5 | 111.6(3) |
| O12 | Mn1 | N28 | 157.38(10) | O3 | C4 | H1 | 109.4(19) |
| O21 | Mn1 | N10 | 111.09(10) | O3 | C4 | H2 | 107(2) |
| O21 | Mn1 | N28 | 80.28(10) | C5 | C4 | H1 | 108(2) |
| N10 | Mn1 | N28 | 89.94(10) | C5 | C4 | H2 | 110(2) |
| CL30 | Mn2 | CL31 | 115.35(4) | H1 | C4 | H2 | 111(3) |
| CL30 | Mn2 | O12 | 125.47(7) | N10 | C5 | C4 | 116.2(3) |
| CL30 | Mn2 | O21 | 94.98(6) | N10 | C5 | C6 | 122.9(3) |
| CL30 | Mn2 | N19 | 107.06(7) | C4 | C5 | C6 | 120.9(3) |
| CL31 | Mn2 | O12 | 118.91(7) | C5 | C6 | C7 | 117.4(3) |
| CL31 | Mn2 | O21 | 103.59(7) | C5 | C6 | H3 | 119.6(17) |
| CL31 | Mn2 | N19 | 95.27(8) | C7 | C6 | H3 | 123.0(17) |
| O12 | Mn2 | O21 | 67.92(8) | C6 | C7 | C8 | 120.0(3) |
| O12 | Mn2 | N19 | 73.41(9) | C6 | C7 | H4 | 118(3) |
| O21 | Mn2 | N19 | 141.32(9) | C8 | C7 | H4 | 122(3) |
| Mn1 | O3 | Mn1 | 102.43(9) | C7 | C8 | C9 | 119.8(3) |
| Mn1 | O3 | C4 | 132.5(2) | C7 | C8 | H5 | 128(2) |
| Mn1 | O3 | C4 | 122.11(19) | C9 | C8 | H5 | 112(2) |
| Mn1 | O12 | Mn2 | 107.75(10) | N10 | C9 | C8 | 120.3(3) |
| Mn1 | O12 | C13 | 128.9(2) | N10 | C9 | C11 | 117.9(3) |
| Mn2 | O12 | C13 | 122.93(19) | C8 | C9 | C11 | 121.8(3) |
| Mn1 | O21 | Mn2 | 106.04(10) | C9 | C11 | H6 | 109(3) |
| Mn1 | O21 | C22 | 119.7(2) | C9 | C11 | H7 | 110(3) |
| C9  | C11 | H8  | 111(2)   | H14 | C20 | H16 | 110(5)   |
| H6  | C11 | H7  | 111(4)   | H15 | C20 | H16 | 114(3)   |
| H6  | C11 | H8  | 104(3)   | O21 | C22 | C23 | 110.8(3) |
| H7  | C11 | H8  | 112(3)   | O21 | C22 | H17 | 108(2)   |
| O12 | C13 | C14 | 110.7(3) | O21 | C22 | H18 | 109(3)   |
| O12 | C13 | H9  | 108(3)   | C23 | C22 | H17 | 109(2)   |
| O12 | C13 | H10 | 110(2)   | C23 | C22 | H18 | 109(3)   |
| C14 | C13 | H9  | 107(2)   | H17 | C22 | H18 | 111(4)   |
| C14 | C13 | H10 | 113(2)   | N28 | C23 | C22 | 116.1(3) |
| H9  | C13 | H10 | 108(3)   | N28 | C23 | C24 | 122.2(3) |
| N19 | C14 | C13 | 117.8(3) | C22 | C23 | C24 | 121.6(3) |
| N19 | C14 | C15 | 122.0(3) | C23 | C24 | C25 | 118.9(4) |
| C13 | C14 | C15 | 120.2(3) | C23 | C24 | H19 | 119(3)   |
| C14 | C15 | C16 | 119.2(4) | C25 | C24 | H19 | 122(3)   |
| C14 | C15 | H11 | 114(3)   | C24 | C25 | C26 | 119.6(4) |
| C16 | C15 | H11 | 127(3)   | C24 | C25 | H20 | 121(3)   |
| C15 | C16 | C17 | 119.2(4) | C26 | C25 | H20 | 119(3)   |
| C15 | C16 | H12 | 120(3)   | C25 | C26 | C27 | 119.9(3) |
| C17 | C16 | H12 | 120(3)   | C25 | C26 | H21 | 122(2)   |
| C16 | C17 | C18 | 120.1(4) | C27 | C26 | H21 | 118(2)   |
| C16 | C17 | H13 | 117(3)   | N28 | C27 | C26 | 120.4(3) |
| C18 | C17 | H13 | 123(3)   | N28 | C27 | C29 | 118.9(3) |
| N19 | C18 | C17 | 120.6(3) | C26 | C27 | C29 | 120.7(3) |
| N19 | C18 | C20 | 117.1(3) | C27 | C29 | H22 | 111(2)   |
| C17 | C18 | C20 | 122.3(4) | C27 | C29 | H23 | 121(2)   |
| C18 | C20 | H14 | 116(4)   | C27 | C29 | H24 | 110(3)   |
| C18 | C20 | H15 | 109(2)   | H22 | C29 | H23 | 106(4)   |
| C18 | C20 | H16 | 113(3)   | H22 | C29 | H24 | 110(4)   |
| H14 | C20 | H15 | 92(4)    | H23 | C29 | H24 | 99(4)    |